\documentclass{sig-alternate}
\usepackage{mathptmx} 

\usepackage{fancyhdr}
\usepackage[normalem]{ulem}
\usepackage[hyphens]{url}
\usepackage[sort,nocompress]{cite}
\usepackage[final]{microtype}
\usepackage[bookmarks=true,breaklinks=true,letterpaper=true,colorlinks,linkcolor=black,citecolor=blue,urlcolor=black]{hyperref}

\usepackage{tabularx}
\usepackage{pifont}
\usepackage{makecell}
\usepackage{multirow}
\usepackage{stfloats}
\usepackage{setspace}
\usepackage[usenames,dvipsnames]{color}
\usepackage[linesnumbered,ruled]{algorithm2e}
\definecolor{dred}{RGB}{191,0,0}
\newfont{\ttlfntuser}{ptmb8t at 15pt}
\fancypagestyle{firstpage}{
  \fancyhf{}
  
  \fancyhead[C]{\vspace{15pt}\normalsize{}} 
  \fancyfoot[C]{\thepage}
}

\fancypagestyle{firstpage}{
  \fancyhf{}
  
  \fancyhead[C]{\vspace{15pt}\normalsize{Sys-Inventor Technical Report; PI: Lei Liu; liulei2010@{buaa.edu.cn/ict.ac.cn}}} 
  \fancyfoot[C]{\thepage}
}

\title{\ttlfntuser Intelligent Resource Scheduling for Co-located Latency-critical Services: \\A Multi-Model Collaborative Learning Approach}

\begin{document}
\maketitle
\thispagestyle{firstpage}
\pagestyle{plain}

\fontsize{10}{11}\selectfont

\noindent\textbf{\centerline{\fontsize{12px}{12px}\selectfont{Abstract}}}\\[-9pt]

\noindent \emph{Latency-critical services have been widely deployed in cloud environments. For cost-efficiency, multiple services are usually co-located on a server. Thus, run-time resource scheduling becomes the pivot for QoS control in these complicated co-location cases. However, the scheduling exploration space enlarges rapidly with the increasing server resources, making the schedulers hardly provide ideal solutions quickly. More importantly, we observe that there are “resource cliffs” in the scheduling exploration space. They affect the exploration efficiency and always lead to severe QoS fluctuations. Resource cliffs cannot be easily avoided in previous schedulers. To address these problems, we propose a novel ML-based intelligent scheduler – OSML. It learns the correlation between architectural hints (e.g., IPC, cache misses, memory footprint, etc.), scheduling solutions and the QoS demands based on a data set we collected from 11 widely deployed services running on off-the-shelf servers. OSML employs multiple ML models to work collaboratively to predict QoS variations, shepherd the scheduling, and recover from QoS violations in complicated co-location cases. OSML can intelligently avoid resource cliffs during scheduling and reach an optimal solution much faster than previous approaches for co-located LC services. Experimental results show that OSML supports higher loads and meets QoS targets with lower scheduling overheads and shorter convergence time than previous studies.
}\\[-1pt]

\noindent\textbf{\fontsize{12px}{12px}\selectfont{1.~~Introduction}}
\\[-3pt]

\textheight=650pt

\noindent Cloud applications are shifting from monolithic architectures to loosely-coupled designs, consisting of many latency-critical (LC) services (e.g., some microservices, interactive services) with strict QoS requirements [18,19,52,53]. Many cloud providers, including Amazon, Alibaba, Facebook, Google, and LinkedIn, employ this loosely-coupled design to improve productivity, scalability, functionality, and reliability of their cloud systems [2,3,18,52]. 

QoS-driven resource scheduling faces more challenges in this era. The cost-efficiency policy drives providers to co-locate as many applications as possible on a server. However, these co-located services exhibit different behaviors across multiple interactive resources, such as CPU cores, cache, bandwidth, and main memory banks. These behaviors also can be drastically different from demand to demand and change within seconds. Moreover, with the increasing number of cores, more threads contend for the shared LLC (last-level cache) and memory bandwidth. Notably, these shared resources interact with each other [32,33,45]. All these issues make resource scheduling for co-located LC services more complicated and time-consuming. Moreover, in reality, end-users keep increasing demands for quick responses from the cloud [15,47,53]. According to Amazon’s estimation, even if the end-users experience a 1-second delay, they tend to give up the transactions, translating to \$1.6 billion loss annually [1]. Briefly, unprecedented challenges are posed for resource scheduling mechanisms [8,10,31,47,52].

Some previous studies [17,31,33,45,61] design clustering approaches to allocate LLC or LLC together with main memory bandwidth to cores for scheduling single-threaded applications. Yet, they are not suitable in cloud environments, as the cloud services often have many concurrent threads and strict QoS constraints (i.e., latency-critical). Alternatively, the existing representative studies either use heuristic algorithms – increasing/decreasing one resource at a time and observing the performance variations [10] or use learning-based algorithms (e.g., Bayesian optimization [46]) in a relatively straightforward manner. The studies in [10,46] show that scheduling five co-located interactive services to meet certain QoS constraints can take more than 20 seconds on average. Existing schedulers still have room for improvement in scheduling convergence time, intelligence, and how to schedule complicated interactive resources (e.g., parallel computing units and complex memory hierarchies) in a timely fashion. Moreover, the existing schedulers can hardly avoid ``resource cliffs'', i.e., decreasing a resource only slightly during scheduling leads to a significant QoS slowdown. To the best of our knowledge, our community has been expecting new directions on developing resource-scheduling mechanisms to handle co-located LC services [16,30,31,45].

To this end, we design OSML, a novel machine learning (ML) based resource scheduler for LC services on large-scale servers. In practice, using ML models significantly improves scheduling exploration efficiency for multiple co-located cloud services and can handle the complicated resource sharing, under/over-provision cases timely. ML has achieved tremendous success in improving speech recognition [54], benefiting image recognition [25], and helping the machine to beat the human champion at Go [13,24,51]. In OSML, we make progress in leveraging ML for resource scheduling, and we make the following contributions.

\noindent\textbf{(1) Investigation in RCliff for Multiple Resources during Scheduling.} We study resource cliff (RCliff, i.e., reducing a resource only slightly leads to a significant QoS slowdown) for computing and cache resources. More importantly, we show that RCliffs commonly exist in many widely used LC services and find the heuristic schedulers can hardly avoid RCliff (Sec.3.3), always leading to a sudden and sharp QoS slowdown. Furthermore, we show ML can be an ideal approach that benefits scheduling (Sec.4.4).

\noindent\textbf{(2) Collaborative ML Models for Intelligent Scheduling.} \textcolor{black}{OSML is an ML-based scheduler that intelligently schedules multiple interactive resources to meet co-located services’ QoS targets. OSML learns the correlation between architectural hints (e.g., IPC, cache misses, memory footprint, etc.), optimal scheduling solutions, and the QoS demands. It employs MLP models to avoid RCliffs intelligently, thus avoiding the sudden QoS slowdown often incurred by the RCliffs in prior schedulers; it predicts the QoS variations and resource margins, and then delivers appropriate resource allocations. It leverages an enhanced DQN to shepherd the allocations and recover from the QoS violation and resource over-provision cases. Moreover, as OSML's models are lightweight and their functions are clearly defined, it is easy to locate the problems and debug them.}

\noindent\textbf{(3) An Open-sourced Data Set for Low-overhead ML.} We have collected the performance traces for widely deployed LC services (in Table \ref{tbl1}), covering \textcolor{black}{62,720,264} resource allocation cases (including the corner cases) that contain around 2-billion samples. These data have a rich set of information, e.g., the RCliffs for multiple resources; the interactions between workload features and the mainstream architectures. Our models can be trained and generalized with these data and then used on new platforms with low-overhead transfer learning. We show our data-driven approach is promising in terms of performance and deployment efficiency. We will make our data set publicly available; people can study it and efficiently train their models \emph{without} an extended period for data collection.

\noindent\textbf{(4) Real Implementation and Detailed Comparisons.} We implement OSML based on latest Linux. OSML is designed as a co-worker of the OS scheduler located between the OS kernel and the user layer. We compare OSML with the most related open-source studies [10,46] and show the advantages.

\begin{table}[!tp]\fontsize{8}{9}\selectfont
\setlength{\tabcolsep}{1pt}
\setlength{\extrarowheight}{0.5pt}
  \centering
    \caption{Latency-critical (LC) services, including micro-/interactive services [18,19,52,68,46]. The max load - max RPS - is with the 99th percentile tail latency QoS target [10,18,46,52].}
  \begin{tabular}{|p{0.22\linewidth}<{\centering}| p{0.28\linewidth}<{\centering} | p{0.46\linewidth}<{\centering} | }
    \hline
    \textbf{LC service} & \textbf{Domain} & \textbf{RPS (Requests Per Second)}\\ 
    \hline
    \hline
	Img-dnn [62]&	Image recognition&	2000,3000,4000,5000,6000 (Max)\\
	\hline
	Masstree [62]&	Key-value store&	3000,3400,3800,4200,4600\\
	\hline
	Memcached [65]&	Key-value store&	256k,512k,768k,1024k,1280k\\
	\hline
	MongoDB [64]&	Persistent database&	1000,3000,5000,7000,9000\\
	\hline
	Moses [62]&	RT translation&	2200,2400,2600,2800,3000\\
	\hline
	Nginx [66]&	Web server&	60k,120k,180k,240k,300k\\
	\hline
	Specjbb [62]&	Java middleware&	7000,9000,11000,13000,15000\\
	\hline
	Sphinx [62]&	Speech recognition&	1,4,8,12,16\\
	\hline
	Xapian [62]&	Online search&	3600,4400,5200,6000,6800\\
	\hline
	Login [68]&	Login&	300,600,900,1200,1500\\
	\hline
	Ads [68,52]&	Online renting ads&	10,100,1000\\
    \hline
  \end{tabular}
  \label{tbl1}
\end{table}

In practice, OSML captures the applications' online behaviors and forwards them to the ML models running on CPU or GPU, and schedules resources accordingly. \textcolor{black}{Compared with [10,46], OSML takes 36\textasciitilde55\% less time to meet the QoS targets and support 10\textasciitilde50\% of higher loads. OSML supports to reclaim over-provided resources to improve efficiency.} Its ML models are with low run-time overheads. We make OSML open source as planned. \\[-1pt]

\textheight=666pt

\noindent\textbf{\fontsize{12px}{12px}\selectfont{2.~~Background and Motivation}}
\\[-3pt]

\noindent The cloud environment has a trend towards a new model [3,18,52], in which cloud applications comprise numerous distributed LC services (i.e., micro/interactive services), such as key-value storing, database serving, and business applications serving [18,19]. Table \ref{tbl1} includes some widely used ones, and they form a significant fraction of cloud applications [18]. These services have different features and resource demands. 

\textcolor{black}{In terms of the datacenter servers, at present, new servers can have an increased number of cores, larger LLC capacity, larger main memory capacity, higher bandwidth, and the resource scheduling exploration space becomes much larger than ever before as a result. Table \ref{tbl2} compares the two typical servers used at different times.} On the one hand, although modern servers can have more cores and memory resources than ever before, they are not fully exploited in today’s cloud environments. For instance, in Google’s datacenter, the CPU utilization is about 45\textasciitilde53\% and memory utilization ranges from 25\textasciitilde77\% during 25 days, while Alibaba’s cluster exhibits a lower and unstable trend, i.e., 18\textasciitilde40\% for CPU and 42\textasciitilde60\% for memory in 12 hours [32,49], indicating that a lot of resources are wasted. On the other hand, the larger resource scheduling exploration space, which consists of more diverse resources, prohibits the schedulers from achieving the optimal solution quickly. Additionally, cloud applications can have dozens of concurrent threads [10,46]. When several cloud applications run on a server, they share and contend resources across multiple resource layers – cores, LLC, memory bandwidth/banks. Previous studies show they may incur severe performance degradation and unpredictable QoS violations, and propose the scheduling approaches at architecture [9,23,44], OS [31,45,50], and user-level [10,37,38]. \emph{\textcolor{black}{Yet, do they perform ideally for scheduling co-located LC services on modern datacenter servers?}}\\[-2pt]

\begin{scriptsize}
\begin{table}[tp]\small
  \setlength{\tabcolsep}{1pt}
  \centering
  \setlength{\extrarowheight}{0.5pt}
  \caption{\textls[-25]{Our platform specification vs. a server used 10 yrs. before.}}
  \begin{tabular}{|p{0.32\linewidth}<{\centering}| p{0.34\linewidth}<{\centering} | p{0.305\linewidth}<{\centering} |}
    \hline
	\textbf{Conf. / Servers}&\textbf{Our Platform}& \textbf{\textls[-20]{Server (10 Years Ago)}}\\
	\hline
	\hline
	\textbf{CPU Model}&\multicolumn{1}{m{0.34\linewidth}<{\centering}|}{Intel Xeon E5-2697 v4} &Intel i7-860\\
	\hline
	\multicolumn{1}{|m{0.32\linewidth}<{\centering}|}{\textbf{\textls[-30]{Logical Processor Cores}}}&\multicolumn{1}{m{0.34\linewidth}<{\centering}|}{36 Cores (18 phy. cores)}&\multicolumn{1}{m{0.305\linewidth}<{\centering}|}{8 Cores (4 phy. cores)}\\
	\hline
	\textbf{Processor Speed}& 2.3GHz&	2.8GHz\\
	\hline
	\multicolumn{1}{|m{0.32\linewidth}<{\centering}|}{\fontsize{7.5}{8} \selectfont\textbf{Main Memory /\quad Channel / BW}}&\multicolumn{1}{m{0.34\linewidth}<{\centering}|}{\fontsize{7.5}{8} \selectfont 256GB, 2400MHz DDR4 / 4 Channels / 76.8GB/s}&\multicolumn{1}{m{0.305\linewidth}<{\centering}|}{\fontsize{7.5}{8} \selectfont 8GB, 1600MHz DDR3 /\quad 2 Channels / 25.6GB/s}\\
	\hline
	\multicolumn{1}{|m{0.32\linewidth}<{\centering}|}{\fontsize{7.5}{8} \selectfont\textbf{Private L1 \& L2 \quad Cache Size}}&32KB and 256KB&32KB and 256KB\\
	\hline
	\textbf{Shared L3 Cache Size}&	45MB - 20 ways&8MB - 16 ways\\
	\hline
	\textbf{Disk}&1TB,7200 RPM,HD&	500GB,5400 RPM,HD\\
	\hline
	\textbf{GPU}&\multicolumn{1}{m{0.34\linewidth}<{\centering}|}{\fontsize{7.5}{8} \selectfont NVIDIA GP104 \quad \quad \quad [GTX 1080], 8GB Memory}& N/A\\
	\hline
  \end{tabular}
  \label{tbl2}
\end{table}
\end{scriptsize}

\begin{figure*}[!t]
\centering
\includegraphics[width=0.99\linewidth]{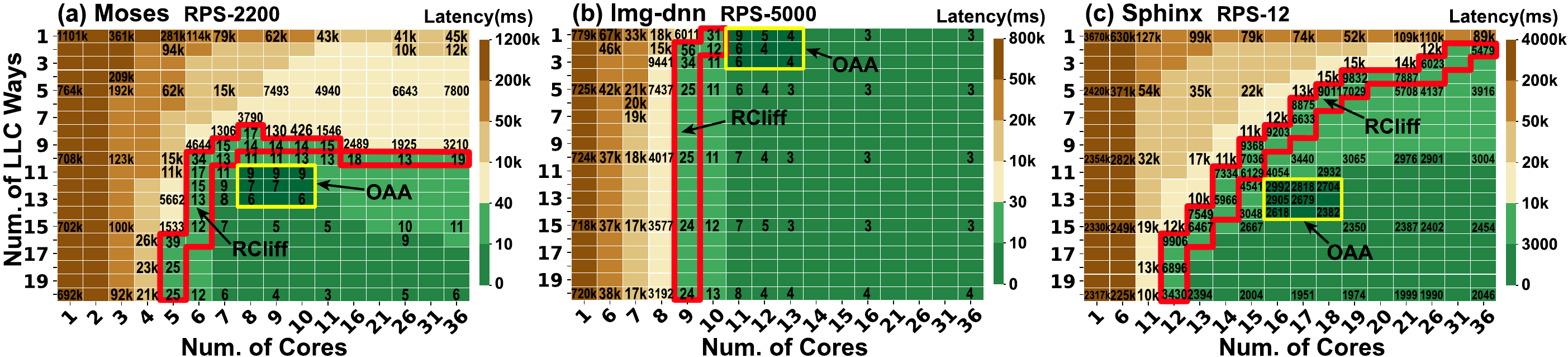}

\caption{The resource scheduling exploration space for cores and LLC ways. All services here are with 36 threads. These figures show the sensitivity to resource allocation under different policies. Each col./row represents a specific number of LLC ways/cores allocated to an application. Each cell denotes the LC service’s response latency under the given number of cores and LLC ways. The Redline highlights the RCliff (can be obtained by selecting the knee solution [69]). The green color cells show allocation policies that bring better performance (low response latency). OAA (Opt Allocation Area) is also illustrated for each LC service. We test all of the LC services in Table \ref{tbl1}. In practice, we find the RCliff and OAA are always existing, though the RPS varies. We only show several of them for the sake of saving space.}
\label{fig1}
\end{figure*}

\noindent\textbf{\fontsize{12px}{12px}\selectfont{3.~~Resource Scheduling for LC Services}}
\\[-3pt]

\noindent To answer the above question, we study the LC services (Table \ref{tbl1}) that are widely deployed in cloud environments.

\begin{spacing}{1.6}
\noindent\textbf{\fontsize{10}{10}\selectfont{3.1.~~Understanding the LC Services - Resource Cliff}}
\end{spacing}

\noindent We study how sensitive these LC services are to the critical resources, e.g., the number of cores and LLC capacity, on a commercial platform (“our platform” in Table \ref{tbl2}). For Moses, as illustrated in Figure \ref{fig1}-a, with the increasing number of cores, more threads are mapped on them simultaneously. Meanwhile, for a specific amount of cores, more LLC ways can benefit performance. Thus, we observe the response latency is low when computing and LLC resources are ample (i.e., below 10ms in the area within green color). The overall trends are also observed from other LC services. 

However, we observe the Cliff phenomenon for these services. In Figure \ref{fig1}-a, in the cases where 6 cores are allocated to Moses, the response latency is increased significantly from 34ms to 4644ms if merely one LLC way is reduced (i.e., from 10 ways to 9 ways). Similar phenomena also happen in cases where computing resources are reduced. \textcolor{black}{As slight resource re-allocations bring a significant performance slowdown, we denote this phenomenon as Resource Cliff (\textbf{RCliff}). It is defined as the resource allocation cases that could incur the most significant performance slowdown if resources (e.g., core, cache) are deprived via a fine-grain way in the scheduling exploration space. Take Moses as an example, on the RCliff (denoted by the red box in Figure \ref{fig1}-a), there would be a sharp performance slowdown if only one core or one LLC way (or both) is deprived.} Figure 1-b and c show RCliffs for Img-dnn and Sphinx, respectively. From another angle, RCliff means that a little bit more resources will bring significant performance improvement. Figure \ref{fig1}-a shows that Moses exhibits RCliff for both core and LLC. Moreover, we test the services in Table \ref{tbl1} across various RPS and find the RCliffs always exist, though the RCliffs vary (8.8\% on average) according to different RPS.

The underlying reason for the cache cliff is locality; for the core cliff, the fundamental reason is on queuing theory - the latency will increase drastically when the request arrival rate exceeds the available cores. \textcolor{black}{RCliff alerts the scheduler not to allocate resources close to it because it is “dangerous to fall off the cliff” and incurs a significant performance slowdown, i.e., even a slight resource reduction can incur a severe slowdown.} Notably, in Figure \ref{fig1}, we highlight each LC service’s \textbf{Optimal Allocation Area (OAA)} in the scheduling exploration space, defined as the ideal number of allocated cores and LLC ways to bring an acceptable QoS. More resources than OAA cannot deliver more significant performance, but fewer resources lead to the danger of falling off the RCliff. OAA is the goal that schedulers should achieve.

\begin{spacing}{1.6}
\noindent\textbf{\fontsize{10}{10}\selectfont{3.2.~~Is OAA Sensitive to the Number of Threads?}}
\end{spacing}

\noindent In practice, an LC service may have many threads for higher performance. Therefore, we come up with the question: \emph{is the OAA sensitive to the number of threads, i.e., if someone starts more threads, will the OAA change?}

\begin{figure}[t]
\vspace{0.1cm}
\centering
\includegraphics[width=0.99\linewidth]{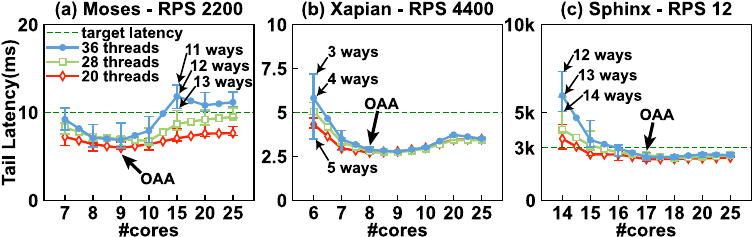}

\caption{\textls[-10]{OAA exists regardless of the num. of concurrent threads.}}
\label{fig2}
\end{figure}

To answer this question, for a specific LC service, we start a different number of threads and map them across a different number of cores (the num. of threads can be larger than the num. of cores). Through the experiments, we observe - (i) More threads do not necessarily bring more benefits. Take Moses as an example, when more threads are started (e.g., 20/28/36) and mapped across a different number of cores, the overall response latency can be higher (as illustrated in Figure \ref{fig2}). The underlying reason lies in more memory contentions at memory hierarchy and more context switch overheads, leading to a higher response latency [20,36]. (ii) The OAA is not sensitive to the number of concurrent threads. For Moses in Figure \ref{fig2}, even if 20/28/36 threads are mapped to 10\textasciitilde25 cores, around 8/9-core cases always perform ideally. Other LC services in Table \ref{tbl1} also show a similar phenomenon, though their OAAs are different for different applications.

In practice, if the QoS for a specific LC service is satisfied, LLC ways should be allocated as less as possible, saving LLC space for other applications. Similarly, we also try to allocate fewer cores for saving computing resources. Here, we conclude that the OAA is not sensitive to the number of threads. We should further reveal: \emph{how do the existing schedulers perform in front of OAAs and RCliffs?}

\textheight=655pt

\begin{spacing}{1.6}
\noindent\textbf{\fontsize{10}{10}\selectfont{3.3.~~Issues the Existing Schedulers May Meet}}
\end{spacing}

\noindent We find three main shortcomings in the existing schedulers when dealing with OAAs and RCliffs. \textbf{(1) Entangling with RCliffs.} \textcolor{black}{RCliffs challenge the schedulers. Many schedulers often employ heuristic scheduling algorithms, i.e., they increase/reduce resources until the monitor alerts that the system performance is suffering a significant change (e.g., a severe slowdown). Yet, these approaches could incur unpredictable latency spiking. For example, if the current resource allocation for an LC service is in the base of RCliff (i.e., the yellow color area in Figure \ref{fig1}-a/b/c), the scheduler has to try to achieve OAA. However, as the scheduler doesn't know the “location” of OAA in the exploration space, it has to increase resources step by step in a fine-grain way, thus the entire scheduling process from the base of the RCliff will incur very high response latency. For another example, if the current resource allocation is on the RCliff or close to RCliff, a slight resource reduction for any purpose could incur a sudden and sharp performance drop for LC services. The previous efforts [10,32,50,53] find there would be about hundreds/thousands of times latency jitter, indicating the QoS cannot be guaranteed during these periods. Thus, RCliffs are essential and should not be neglected when designing a scheduler.} \textbf{(2) Unable to accurately and simultaneously schedule a combination of multiple interactive resources (e.g., core counts, LLC ways and bandwidth usage) to achieve OAAs in low overheads.} Prior studies [10,31,32,45] show that the core computing ability, cache hierarchy, and memory bandwidth are interactive factors for resource scheduling. Solely considering a single dimension in scheduling often leads to sub-optimal QoS. However, the existing schedulers using heuristic or model-based algorithms usually schedule one dimension resource at a time and bring high overheads on scheduling multiple interactive resources. For example, the state-of-the-art work PARTIES [10] takes around 20\textasciitilde30 seconds on average (up to 60 seconds in the worst cases) to find ideal allocations when 3\textasciitilde6 LC services are co-running. The efforts in [16,41,42] also show the heuristics inefficiency due to the high overheads on scheduling various resources with complex configurations. \textbf{(3) Unable to provide accurate QoS predictions.} Therefore, the scheduler can hardly balance the global QoS and resource allocations across all co-located applications, leading to QoS violations or resource over-provision.

\textls[-14]{An ideal scheduler should avoid the RCliff and quickly achieve the OAA from any positions in the scheduling space. We claim it is time to design a new scheduler, and using ML can be a good approach to handle such complicated cases in low overheads.}\\[-1pt]

\begin{scriptsize}
\begin{table}[t]\small
  \setlength{\tabcolsep}{1pt}
  \centering
  \setlength{\extrarowheight}{0.5pt}
  \caption{The input parameters for ML models.}
  \begin{tabular}{|p{0.24\linewidth}<{\centering}| p{0.50\linewidth}<{\centering} | p{0.22\linewidth}<{\centering} |}
    \hline
    \textbf{Feature} & \textbf{Description} & \textbf{Models}\\ 
    \hline
    \hline
	IPC&	Instructions per clock&	A/A’/B/B’/C\\
	\hline
	Cache Misses&	LLC misses per second&	A/A’/B/B’/C\\
	\hline
	MBL&	Local memory bandwidth&	A/A’/B/B’/C\\
	\hline
	CPU Usage&	The sum of each core’s utilization&	A/A’/B/B’/C\\
	\hline
	Virt. Memory&	Virtual memory in use by an app&	A/A’/B/B’\\
	\hline
	Res. Memory&	Resident memory in use by an app&	A/A’/B/B’\\
	\hline
	Allocated Cores&	The number of allocated cores&	A/A’/B/B’/C\\
	\hline
	Allocated Cache&	The capacity of allocated cache& 	A/A’/B/B’/C\\
	\hline
	\textcolor{black}{Core Frequency}&	\textcolor{black}{Core Frequency during run time}&	\textcolor{black}{A/A’/B/B’/C}\\
	\hline
	QoS Slowdown&	Percentage of QoS slowdown&	B\\
	\hline
	Expected Cores&	Expected cores after deprivation&	B’\\
	\hline
	Expected Cache&	 Expected cache after deprivation&	B’\\
	\hline
	Cores used by N.&	Cores used by Neighbors&	A’/B/B’\\
	\hline
	Cache used by N.&	 Cache capacity used by Neighbors&	A’/B/B’\\
	\hline
	MBL used by N.&	Memory BW used by Neighbors&	A’/B/B’\\
	\hline
	Resp. Latency&	Average latency of a LC service&	C\\
	\hline
  \end{tabular}
  \label{tbl3}
\end{table}
\end{scriptsize}

\noindent\textbf{\fontsize{12px}{12px}\selectfont{4.~~Leveraging ML for Scheduling}}
\\[-3pt]

\noindent We build a new resource scheduler – OSML, which differs from the previous schedulers in several ways. (1) It uses data-driven static ML models and reinforcement learning model to work collaboratively to perform scheduling. Model-A predicts the OAA and the RCliff for a specific LC service. Model-B balances the QoS and resource allocations among co-located LC services. Model-C is an online reinforcement learning model that dynamically shepherds the allocations. (2) We collect extensive real traces for widely deployed LC services, making using data-driven ML technologies practical in cloud systems and providing more accurate predictions as a result.

\begin{spacing}{1.6}
\noindent\textbf{\fontsize{10}{10}\selectfont{4.1.~~Model-A: Aiming OAA}}
\end{spacing}

\noindent \textbf{Model-A Description.} The neural network used in Model-A is a 3-layer multi-layer perceptron (MLP); each layer is a set of nonlinear functions of a weighted sum of all outputs that are fully connected from the prior one [21,24]. There are 40 neurons in each hidden layer. There is a dropout layer with a loss rate of 30\% behind each fully connected layer to prevent overfitting. For each LC service, the \textbf{inputs} of the MLP include 9 items in Table \ref{tbl3} and the \textbf{outputs} include the OAA for multiple interactive resources, OAA bandwidth (bandwidth requirement for OAA), and the RCliff. Model-A has a shadow – A’, which has the same MLP structure and 12 input parameters (Table \ref{tbl3}), providing solutions when multiple LC services are running together.

\textbf{Model-A Training.} Collecting training data is an offline job. We have collected the performance traces that involve the parameters in Table \ref{tbl3} for the LC services in Table \ref{tbl1} on "our platform" in Table \ref{tbl2}. The parameters are normalized into [0,1] according to the function: $\rm Normalized\_Feature=(Feature-Min)/(Max-Min)$. Feature is the original value; Max and Min are predefined according to different metrics. 

\begin{figure}[t]
	\centering
	\begin{minipage}[t]{0.52\linewidth}
		\centering
		\includegraphics[width=0.80\linewidth]{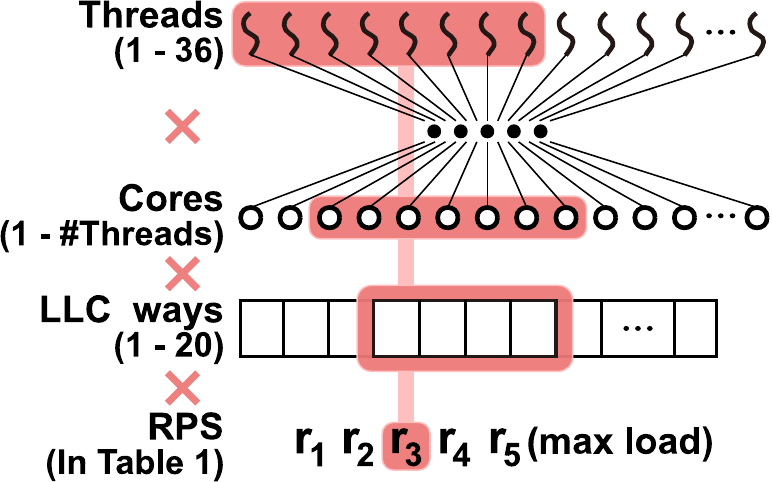}
		\vspace{2pt}
		\caption{\textls[-20]{Model-A data collection.}}
		\label{fig3}
	\end{minipage}
	\begin{minipage}[t]{0.47\linewidth}
		\flushright
		\includegraphics[width=0.99\linewidth]{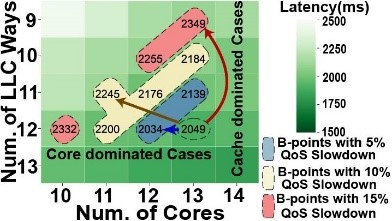}
		\caption{Model-B training.}
		\label{fig4}
	\end{minipage}
\end{figure}

For each LC service with every common RPS demand, we sweep 36 threads to 1 thread across LLC allocation policies ranging from 1 to 20 ways and map the threads on a certain number of cores and collect the performance trace data accordingly. In each case, we label the corresponding OAA, RCliff and OAA bandwidth. For example, Figure \ref{fig3} shows a data collection case where 8 threads are mapped onto 7 cores with 4 LLC ways. We feed the LC services with diverse RPS (Table \ref{tbl1}), covering most of the common cases. \textcolor{black}{Moreover, to train Model-A's shadow (A'), we map LC services on the remaining resources in the above process and get the traces for co-location cases. We test all co-location cases for LC services in Table \ref{tbl1}, and find the LC services' RCliffs vary from 2.8\% to 38.9\%, and OAAs vary from 5.6\% to 36.1\%, respectively, when multiple LC services are co-running. Our dataset contains more details.}

Finally, we collect \textcolor{black}{43,299,135} data tuples, covering \textcolor{black}{1,308,296} allocation cases with different numbers of cores, LLC ways, and bandwidth allocations. A large amount of traces lead to higher accuracy for ML models. The workload characteristics are converted to comprehensive traces consisting of diverse hardware parameters used for fitting and training MLP to provide predictions.

\begin{spacing}{1.5}
\noindent\textbf{\fontsize{10}{10}\selectfont{4.2.~~Model-B: Balancing QoS and Resource Allocations}}
\end{spacing}

\noindent \textbf{Model-B Description.} Model-B employs an MLP with the same structure in Model-A’ plus one more input item, i.e., QoS slowdown (Table \ref{tbl3}). Model-B \textbf{outputs} the minimum resources that a service can be deprived of under allowable QoS slowdown. As the computing units and memory resource can be fungible [10], Model-B’s outputs include three policies, i.e., <cores, LLC ways>, <cores dominated, LLC ways> and <cores, LLC ways dominated>, respectively. The tuple items are the number of cores, LLC ways deprived and reallocated to others with the allowable QoS slowdown. The term “cores dominated” indicates the policy using more cores to trade the LLC ways, and vice versa. The allowable QoS slowdown is determined according to the user requirement or the LC services’ priority and controlled by the OSML's central logic. We denote Model-B's outputs as B-Points. 

Model-B trades QoS for resources. For example, when an LC service (E) comes to a server that already has 4 co-located services, OSML enables Model-A' to obtain <n+, m+>, which denotes at least n more cores and m more LLC ways should be provided to meet E’s QoS. Then, OSML enables Model-B and uses the allowable QoS slowdown as an input to infer B-Points for obtaining resources from other co-located services. B-Points include the “can be deprived” resources from E's neighbors with the allowable QoS slowdown. Finally, OSML finds the best solution to match <n+, m+> with B-Points, which has a minimal impact on the co-located applications' current allocation state. Detailed logic is in Algo.\_1. Besides, we design Model-B’ (a shadow of Model-B) to predict how much QoS slowdown will suffer if a certain amount of resources is deprived of a specific service. The structure of Model-B’ is similar to Model-B.

\textbf{Model-B Training.} For training Model-B and B’, we reduce the allocated resources for a specific LC service from its OAA by fine-grain approaches, as illustrated in Figure \ref{fig4}. The reduction has three angles, i.e., horizontal, oblique, and vertical, i.e., B-Points include <cores dominated, LLC ways>, <cores, LLC ways>, <cores, LLC ways dominated>, respectively. For each fine-grain resource reduction step, we collect the corresponding QoS slowdowns and then label them as less equal to ($\le$) 5\%, 10\%, 15\%, and so on, respectively. Examples are illustrated in Figure \ref{fig4}, which shows the B-Points with the corresponding QoS slowdown. We collect the training data set for every LC service in Table \ref{tbl1}. The data set contains \textcolor{black}{65,998,227} data tuples, covering \textcolor{black}{549,987} cases.

\textbf{Model-B Function.} We design a new loss function:
\begin{equation*} \label{eqn1}
\setlength{\abovedisplayskip}{0pt}
\setlength{\belowdisplayskip}{0pt}
\fontsize{8.5}{8}
\begin{aligned}
\rm{
  L=\frac{1}{n}\sum_{t=1}^{n}\left(\frac{y_t}{y_t+c}\times(s_t-y_t)\right)^2,
}
\end{aligned}
\end{equation*}
in which $\rm s_t$ is the prediction output value of Model-B, $\rm y_t$ is the labeled value in practice, and `c' is a constant that is infinitely close to zero. We multiply the difference between $\rm s_t$ and $\rm y_t$ by $\rm \frac{y_t}{y_t+c}$ for avoiding adjusting the weights during backpropagation in the cases where $\rm y_t=0$ and $\rm \frac{y_t}{y_t+c}=0$ caused by some non-existent cases (we label the non-existent cases as 0, i.e., $\rm y_t=0$, indicating we don’t find a resource-QoS trading policy in the data collection process).  

\begin{spacing}{1.6}
\noindent\textbf{\fontsize{10}{10}\selectfont{4.3.~~Model-C: Handling the Changes On the Fly}}
\end{spacing}

\begin{figure}[t]
\centering
\includegraphics[width=0.99\linewidth]{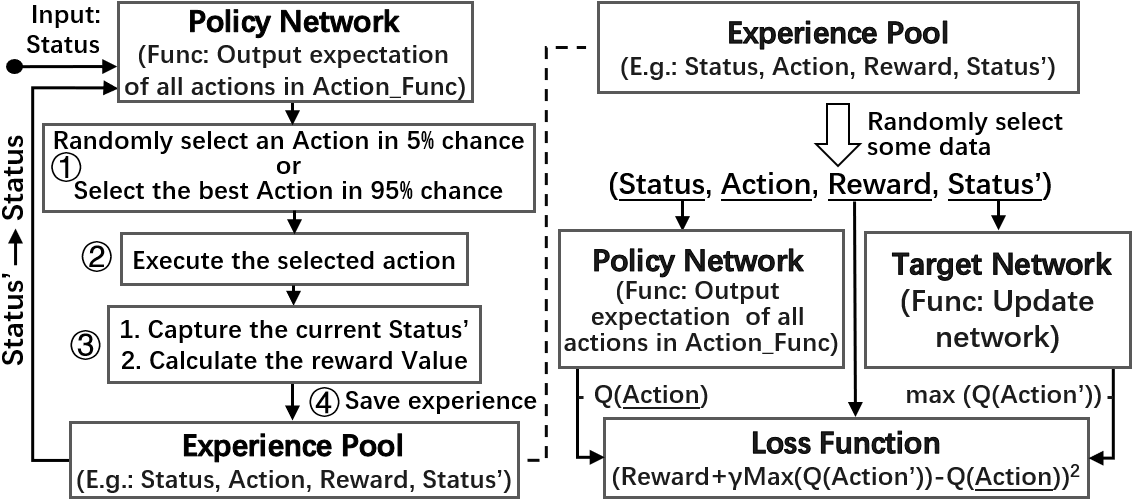}

\caption{Model-C in a nutshell.}
\label{fig5}
\end{figure}

\noindent \textbf{Model-C Description.} Model-C corrects the resource under-/over-provision and conducts online training. Figure \ref{fig5} shows the Model-C in a nutshell. Model-C's core component is an enhanced Deep Q-Network (DQN) [43], consisting of two neural networks, i.e., Policy Network and Target Network. The Policy and Target Network employ the 3-layer MLP, and each hidden layer has 30 neurons. Policy Network’s \textbf{inputs} consist of the parameters in Table \ref{tbl3}, and the \textbf{outputs} are resource scheduling actions (e.g., reducing/increasing a specific number of cores or LLC ways) and the corresponding expectations (defined as Q(action)). These actions numbered 0\textasciitilde48 are defined as Action\_Function:\{$\rm {<m,n>|m\in[-3,3],n \in[-3,3]}$\}, in which a positive m denotes allocating m more cores (i.e., add operation) for an application and a negative m means depriving it of m cores (i.e., sub operation); n indicates the actions on LLC ways. Figure \ref{fig5} illustrates Model-C's logic. The scheduling action with the maximum expectation value (i.e., the action towards the best solution) will be selected in {\fontsize{11}{8}\ding{172}} and executed in {\fontsize{11}{8}\ding{173}}. In {\fontsize{11}{8}\ding{174}}, Model-C will get the Reward value according to the Reward Function. Then, the tuple <Status, Action, Reward, Status’> will be saved in the Experience Pool in {\fontsize{11}{8}\ding{175}}, which will be used during online training. The terms Status and Status’ denote system’s status described by the parameters in Table \ref{tbl3} before and after the Action is taken. Model-C can quickly have the ideal solutions in practice (about 2 or 3 actions). Please note that in {\fontsize{11}{8}\ding{172}}, Model-C might randomly select an Action instead of the best Action with a 5\% chance. By doing so, OSML avoids falling into a local optimum [43].

\textbf{Model-C’s Reward Function.} The reward function of Model-C is defined as follow:
\begin{equation*}
\setlength{\abovedisplayskip}{0pt}
\setlength{\belowdisplayskip}{0pt}
\fontsize{8.5}{8}
\label{eq2}
\begin{aligned}
\rm{If}\ &\rm{Latency_{t-1}}>\rm{Latency_t:}\\[-0.15cm]
\ &\rm{R_t}=\rm{\log (1+Latency_{t-1}-Latency_t)-(\Delta CoreNum + \Delta CacheWay)}\\[-0.15cm]
\rm{If}\ &\rm{Latency_{t-1} < Latency_t:}\\[-0.15cm]
\ &\rm{R_t=-\log (1+Latency_t-Latency_{t-1})-(\Delta CoreNum + \Delta CacheWay)}\\[-0.15cm]
\rm{If}\ &\rm{Latency_{t-1} = Latency_t:}\\[-0.15cm]
\ &\rm{R_{t}=-(\Delta CoreNum + \Delta CacheWay),}
\end{aligned}
\end{equation*}
where $\rm Latency_{t-1}$ and $\rm Latency_t$ denote the latency in previous and current status, respectively; $\rm\Delta CoreNum$ and $\rm\Delta CacheWay$ represent the changes in the number of cores and LLC ways, respectively. This function gives higher reward and expectation to Action that brings less resource usage and lower latency. Thus, Model-C can allocate appropriate resources. Algo.\_2 and 3 show the logic on using Model-C in detail.

\textbf{Offline Training.} A training data tuple includes Status, Status’, Action and Reward, which denote the current status of a LC service, the status after these actions are conducted (e.g., reduce several cores or allocate more LLC ways) and the reward calculated using the above functions, respectively. 

To create the training data set for Model-C, we resort to the data set used in Model-A training. The process is as follows. Two tuples in Model-A training data set are selected to denote Status and Status’, and we further get the differences of the resource allocations between the two status (i.e., the actions that cause the status shifting). Then, we use the reward function to have the reward accordingly. These 4 values form a specific tuple in Model-C training data set. In practice, as there are a large number of data tuples in Model-A training data set, it is impossible to try every pair of tuples in the data set, we only select two tuples from resource allocation policies that have less than or equal to 3 cores, or 3 LLC ways differences. Moreover, we also collect the training data in the cases where LLC sharing occurs among different LC services and save them in the Experience Pool. Using them, Model-C can have the knowledge on selecting actions in resource sharing cases. To sum up, we have \textcolor{black}{1,521,549,190} tuples in Model-C training data set. 

\textbf{Online Training.} Model-C collects online traces. The training flow is in the right part of Figure \ref{fig5}. Model-C randomly selects some data tuples (200 by default) from the Experience Pool. For each tuple, Model-C uses the Policy Network to get the Action’s expectation value (i.e., $\rm Q(Action)$ [43]) with the Status. \textcolor{black}{In Model-C, the target of the Action's expectation value is the Reward observed plus the weighted best expectation value of the next status (i.e., Status’).} \textcolor{black}{As illustrated in Figure 5, Model-C uses the Target Network to have the expectation values of Status’ for the actions in Action\_Function and then finds the best one, i.e., $\rm Max(Q(Action’))$. We design a new Loss Function based on MSE: $\rm(Reward + \gamma Max(Q(Action’)) - Q(Action))^2$. \textcolor{black}{It helps the Policy Network predict closer to the target. The Policy Network is updated during online training. The Target Network's weights are synchronized periodically with the Policy Network's weights. Doing so enables the Target Network to provide stable predictions for the best expectation value of Status’ within a predefined number of time steps, thus improving the stability of the training and prediction.}}

\begin{scriptsize}
\begin{table}[t]\small
  \setlength{\tabcolsep}{1pt}
  \centering
  \setlength{\extrarowheight}{1pt}
  \caption{The Summary of ML models in OSML.}
  \begin{tabular}{|m{0.08\linewidth}<{\centering}| m{0.11\linewidth}<{\centering} | m{0.12\linewidth}<{\centering} |m{0.12\linewidth}<{\centering} | m{0.20\linewidth}<{\centering} | m{0.14\linewidth}<{\centering}|m{0.15\linewidth}<{\centering}|}
    \hline
    \textbf{ML}& \textbf{Model}&\textbf{Features}&\multicolumn{1}{m{0.12\linewidth}<{\centering}|}{\fontsize{7.5}{8}\selectfont\textbf{Model Size}}&\multicolumn{1}{m{0.20\linewidth}<{\centering}|}{\fontsize{7.5}{8}\selectfont\textbf{Loss Function}}&\multicolumn{1}{m{0.14\linewidth}<{\centering}|}{\fontsize{7.5}{8}\selectfont\textbf{Gradient Descent}}&\multicolumn{1}{m{0.15\linewidth}<{\centering}|}{\fontsize{7.5}{8}\selectfont\textbf{Activation Function}}\\
    \hline
    \hline
	A&	MLP&	9&	144 KB&\multirow{2}*{\makecell[c]{Mean Square\\Error (MSE)}}&\multirow{4}*{\makecell[c]{Adam\\Optimizer}}&\multirow{5}*{ReLU}\\
    \cline{1-4}
	A’&	MLP&	12&	155 KB& ~&~&~\\
    \cline{1-5}
	B&	MLP&	13&	110 KB&Modified MSE&	~&~\\
    \cline{1-5}
	B’&	MLP&	14&	106 KB&MSE&	~&~\\
    \cline{1-6}
	C&	DQN&	8&	141 KB&Modified MSE&	RMSProp&~\\
    \hline
  \end{tabular}
  \label{tbl4}
\end{table}
\end{scriptsize}

\begin{spacing}{1.6}
\noindent\textbf{\fontsize{10}{10}\selectfont{4.4.~~Discussions on the design of ML Models}}
\end{spacing}

\noindent \textls[-10]{\textbf{(1) Why using MLPs.} Table \ref{tbl4} characterizes the ML models used in OSML. We employ three-layered MLPs in Model-A and B, because they can fit continuous functions with an arbitrary precision given a sufficient number of neurons in each layer [67], and we can use extensive training data to improve the accuracy of MLPs for predicting OAAs and RCliffs. Moreover, after offline training, using MLPs brings negligible run-time overheads to OSML. \textbf{(2) Why do we have the three models?} We divide the OSML's scheduling logic into three parts, which the three models cover, respectively. Models work in different scheduling phases, and no single model can handle all cases. Model-A predicts the RCliffs and OAAs; Model-B predicts the QoS variations and resource margins in co-location cases. DQN in Model-C learns online to shepherd the scheduling results from Model-A/B. They are necessary and work cooperatively to cover main scheduling cases. Moreover, they are easier to generalize than other approaches, e.g., a table lookup approach (Sec.6.4). \emph{Why not only use the online learning Model-C?} Model-C uses DQN that depends on the start points. It starts with Model-A/B's outputs to avoid exploring the whole (large) scheduling space. Without the approximate OAA provided by Model-A for many unseen cases, only using Model-C will incur more scheduling actions (overheads). \textcolor{black}{\textbf{(3) Generalization for Unseen apps and New servers.} (\romannumeral1) We use ``hold-out cross validation'', i.e., the training data (70\% of the whole data set) excludes the testing data (30\%) for each LC service. (\romannumeral2) We train models with extensive representative traces from many types of LC services (e.g., memory/CPU intensive, random memory access, etc. [10,46]), fitting the correlation between architectural hints (IPC, cache miss, memory footprint, CPU/LLC utilization and MBL are more critical parameters for affecting models' performance), OAA, and the QoS demands. For instance, the spearman correlation coefficient is 0.571, 0.499, and -0.457 between OAA and cache miss, MBL, and IPC, respectively. \emph{On different platforms or for new/unseen applications}, these numbers might be varied; however, this correlation trend is not changed, enabling OSML to generalize to other situations. (\romannumeral3) Using transfer learning, collecting new traces on a new server for several hours will make OSML work well for it (refer to Sec.6.4). (\romannumeral4) OSML is a long-term project open to the community; we continue adding new traces collected from new applications and new servers to the data set for enhancing models’ performance for new cases.}} \\[-1pt]

\textheight=660pt

\begin{figure}[t]
\centering
\includegraphics[width=0.99\linewidth]{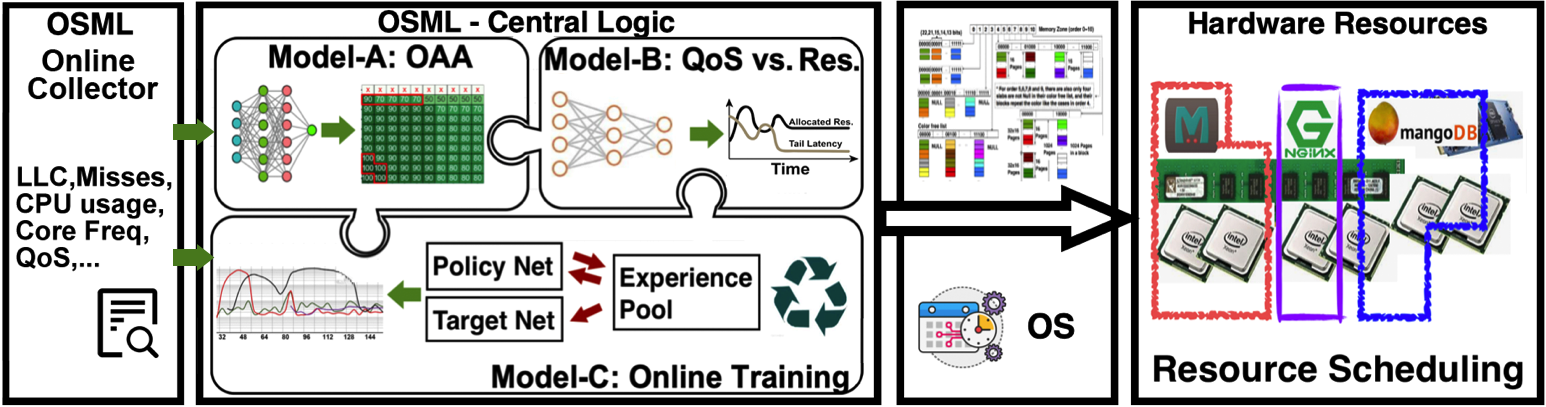}
\caption{The overview of the system design of OSML.}
\label{fig6}
\end{figure}

\begin{figure}[b]
\vspace{0.1cm}
\centering
\includegraphics[width=0.99\linewidth]{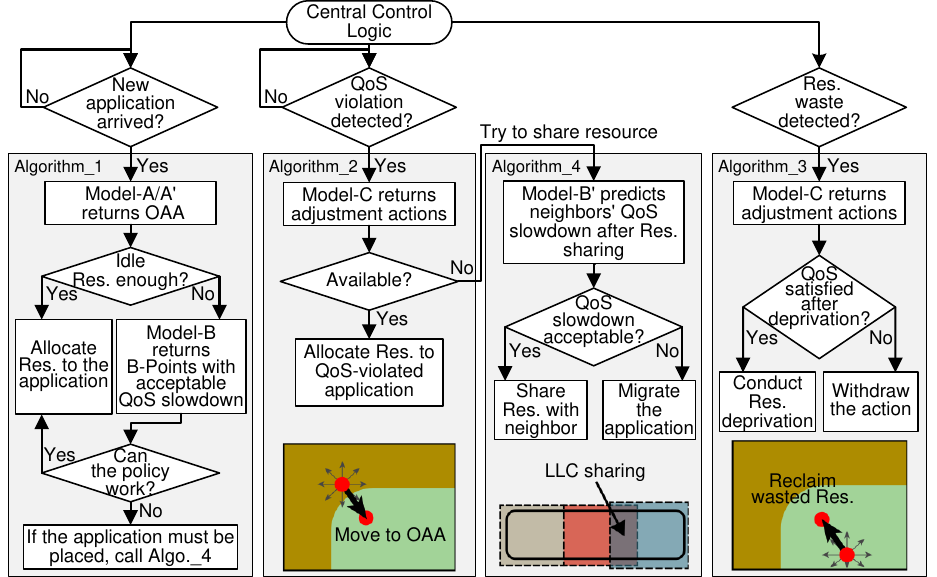}
\caption{OSML’s central logic.}
\label{fig7}
\end{figure}

\noindent\textbf{\fontsize{12px}{12px}\selectfont{5.~~OSML: System Design}}
\begin{spacing}{1.6}
\noindent\textbf{\fontsize{10}{10}\selectfont{5.1.~~The Central Control Logic}}
\end{spacing}

\noindent The overview of the system design of OSML is in Figure \ref{fig6}. The central controller of OSML coordinates the ML models, manages the data/control flow, and reports the scheduling results to the upper scheduler. Figure \ref{fig7} shows its overall control logic. More details are as below.

\textbf{Allocating Resources for LC services.}  Algo.\_1 shows how OSML uses Model-A and B in practice. Figure \ref{fig7} highlights its operations. For a newly coming LC service, the central controller calls Model-A via the interface \emph{modelA\_oaa\_rcliff()} to get the OAA and RCliff. Suppose the current idle resources are not sufficient to satisfy the new LC service. In that case, OSML will enable Model-B through the interface \emph{modelB\_trade\_qos\_res()} to deprive some resources of other LC services with the allowable QoS slowdown (controlled by the upper-level scheduler) and then allocate them to the new one. In the depriving process for a specific LC service, OSML reduces its allocated resources and gets close to the RCliff, but it will not easily fall off the RCliff unless expressly permitted (refer to Algo.\_4).

\textls[-10]{\textbf{Dynamic Adjusting.} Figure \ref{fig7} shows the dynamic adjusting of Algo.\_2 and 3, in which Model-C works as a dominant role. During the run time, OSML monitors each LC service’s QoS status for every second. If the QoS violation is detected, the central controller will enable Algo.\_2 and call Model-C to allocate more resources to achieve the ideal QoS. The interface is \emph{modelC\_upsize()}. If OSML finds an LC service is over-provisioned (i.e., wasting resources), Algo.\_3 will be used to reclaim them, and Model-C will be called via \emph{modelC\_downsize()}.}

\begin{figure}[!t]
\centering
\includegraphics[width=\linewidth]{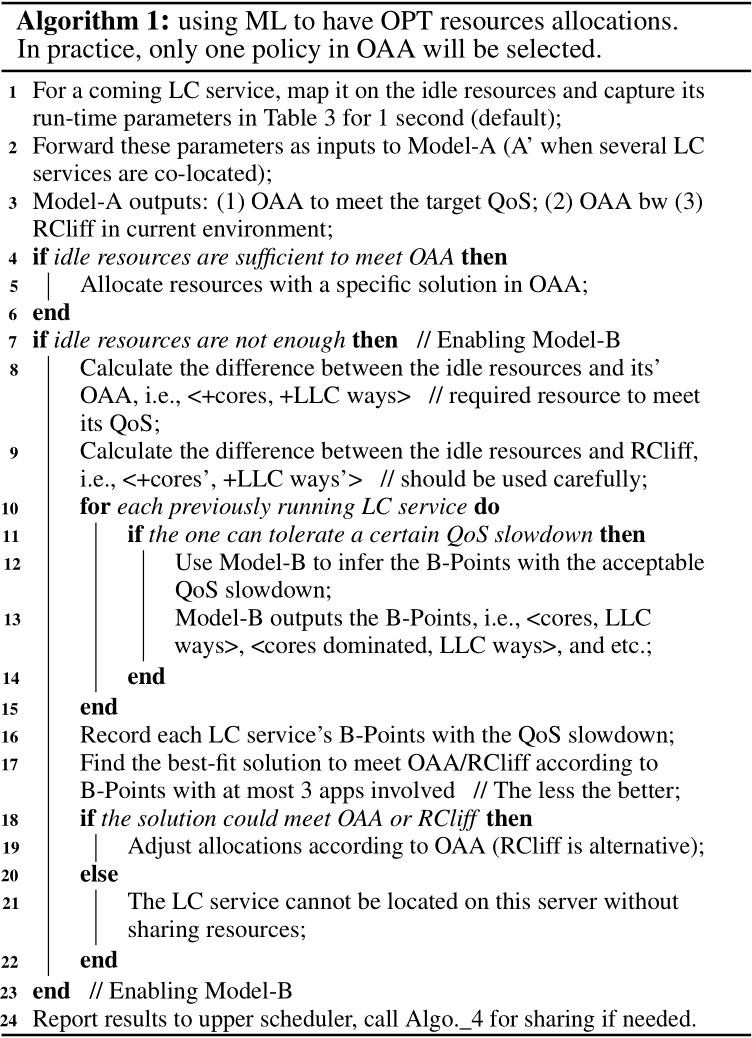}
\label{algo_1}
\vspace{-0.40cm}
\end{figure}

\textls[-5]{Moreover, Algo.\_4 will enable resource sharing (the default scheduling is to do hard partitioning of cores/LLC ways), if all of the co-located LC services are close to their RCliff and the upper scheduler still wants to increase loads on this server. Model-A and B work cooperatively to accomplish this goal in Algo.\_4. In practice, to minimize the adverse effects, resource sharing usually happens between only two applications. Note that Algo.\_4 might incur the resource sharing over the RCliff, and thus may bring higher response latency for one or more LC services. OSML will report the potential QoS slowdown to the upper scheduler and ask for the decisions. If the slowdown is not allowed, the corresponding actions will not be conducted.}

\textbf{Bandwidth Scheduling.} OSML partitions the overall bandwidth for each co-located LC service according to the ratio $\rm BW_j/\sum{BW_i}$. $\rm BW_j$ is a LC service’s OAA bandwidth requirement, which is obtained from the Model-A. Note that such scheduling needs MBA support [4,5] in CPU.

\begin{figure}[t]
\centering
\includegraphics[width=\linewidth]{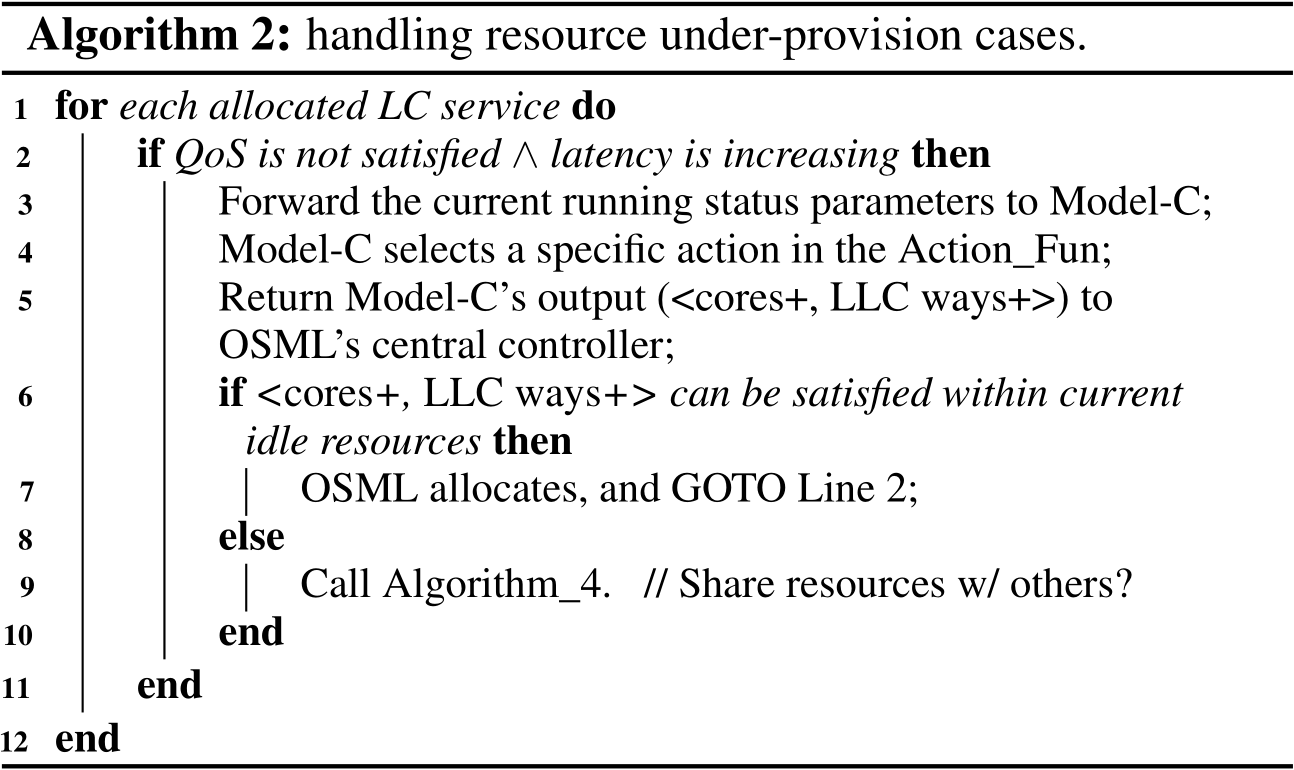}
\label{algo_2}
\vspace{-0.32cm}
\end{figure}
\begin{figure}[t]
\centering
\includegraphics[width=\linewidth]{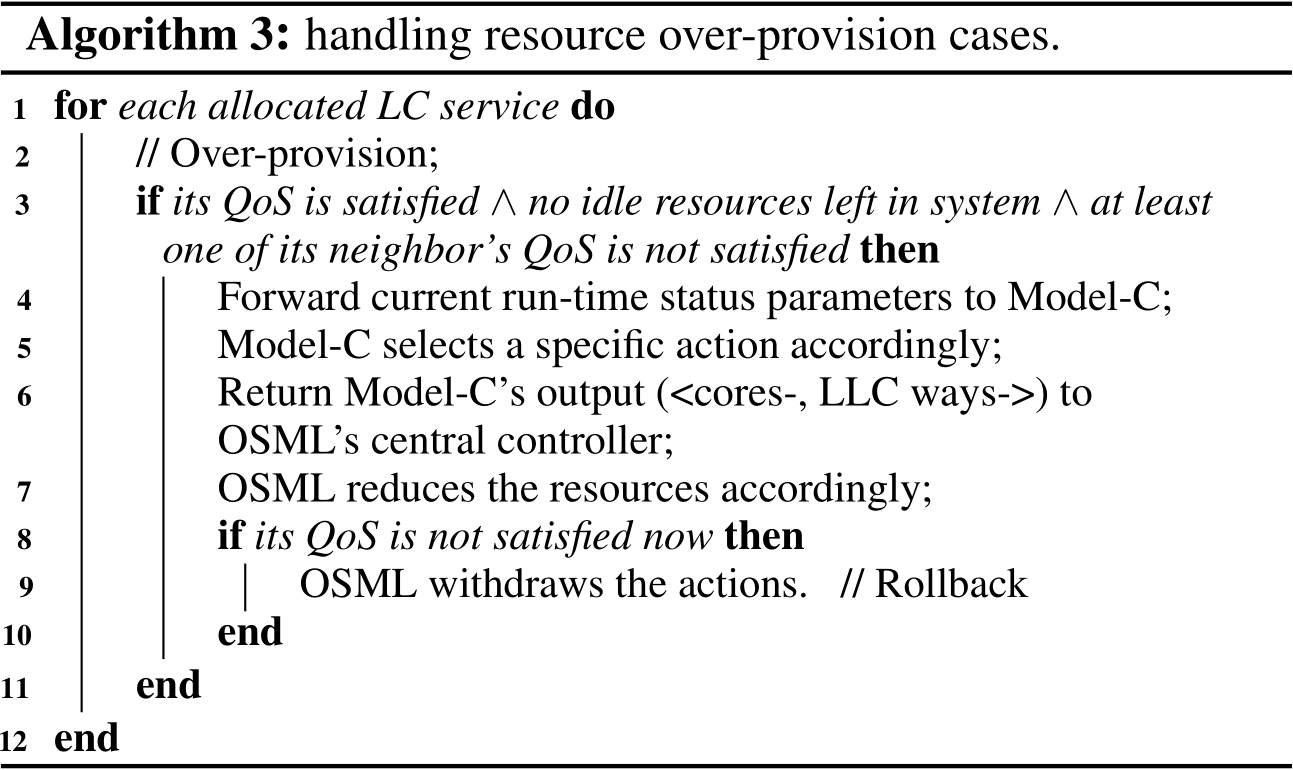}
\label{algo_3}
\vspace{-0.32cm}
\end{figure}
\begin{figure}[t!]
\centering
\includegraphics[width=\linewidth]{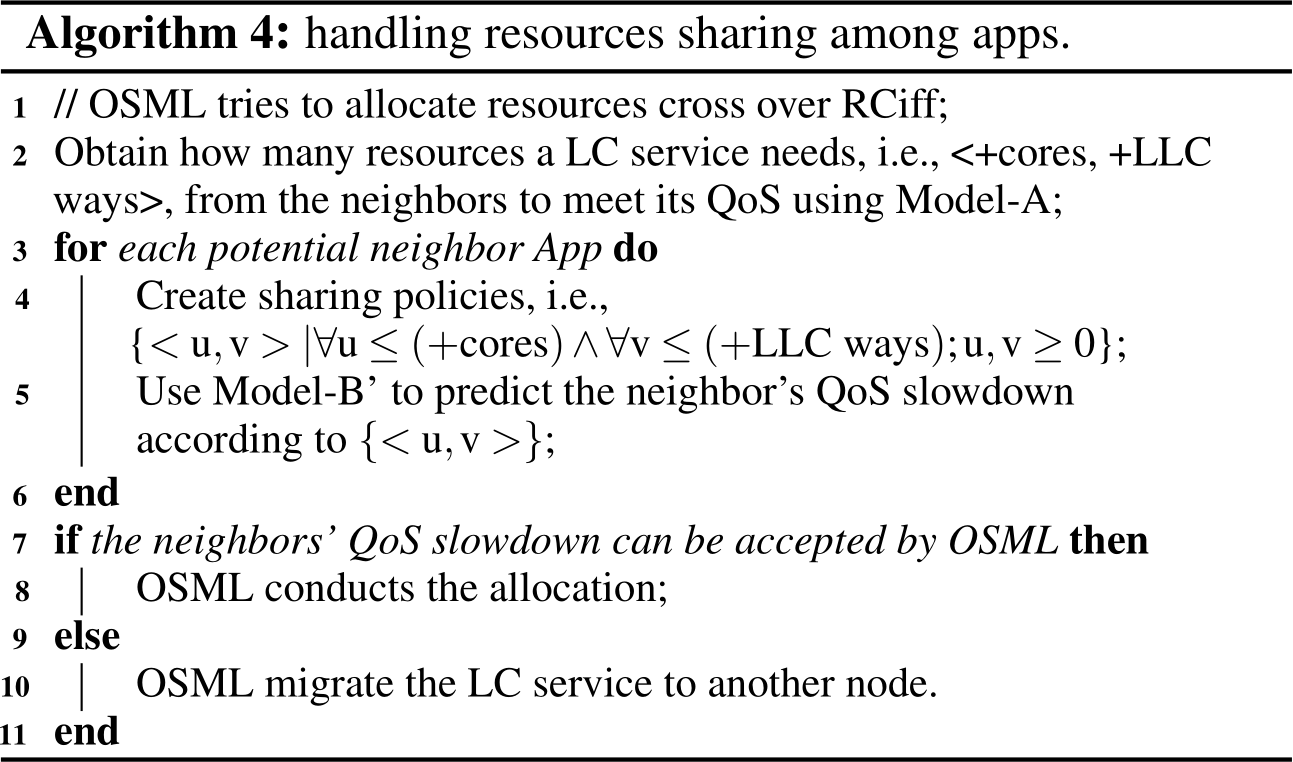}
\label{algo_4}
\vspace{-0.38cm}
\end{figure}

\begin{spacing}{1.6}
\noindent\textbf{\fontsize{10}{10}\selectfont{5.2.~~Implementation}}
\end{spacing}

\noindent OSML interacts with the upper scheduler or clients to obtain the latency of all requests and checks whether they have met their QoS targets. \textcolor{black}{OSML monitors the run-time parameters for each co-located LC service using performance counters for every second (default). If the observation period is too short, other factors (e.g., cache data evicted from the cache hierarchy, context switch) may interfere with the sampling results. Moreover, we find OSML performs well with other interval settings and allows configuration flexibility as needed (e.g., 1.5 or 2 seconds).}

We design OSML that works cooperatively with OS (Figure \ref{fig6}). As the kernel space lacks the support of ML libraries, OSML lies in the user space and exchanges information with the OS kernel. OSML is implemented using python and C. It employs Intel CAT technology [4] to control the cache way allocations, and it supports dynamically adjusting. OSML uses Linux’s taskset and MBA [5] to allocate specific cores and bandwidth to an LC service. OSML captures the online performance parameters by using the pqos tool [4] and PMU [5]. The ML models are based on TensorFlow [6] with the version \textcolor{black}{2.0.4}, and can be run on either CPU or GPU.\\[-1pt]

\textheight=666pt

\noindent\textbf{\fontsize{12px}{12px}\selectfont{6~~Evaluations}}
\begin{spacing}{1.6}
\noindent\textbf{\fontsize{10}{10}\selectfont{6.1.~~Methodology}}
\end{spacing}
\noindent \textls[-8]{\textcolor{black}{We evaluate OSML on our platform in Table \ref{tbl2}. Details on LC applications can be found in Table 1. The metrics involve the QoS (similar to [10], the QoS target of each application is the 99th percentile latency of the knee of the latency-RPS curve. Latency higher than the QoS target is a violation.); Effective Machine Utilization (EMU) [10] (the max aggregated load of all co-located LC services) – higher is better. We first evaluate the scenarios where LC services run at constant loads, and the loads are from 10\% - 100\%. Then, we explore workload churn. We inject applications with loads from 20\% - 100\% of their respective max load. Furthermore, to evaluate the generalization of OSML, we employ some new/unseen applications that are not in Table 1 and the new platform in our experiments. If an allocation in which all applications meet their QoS cannot be found after 3 mins, we signal that the scheduler cannot deliver QoS for that configuration.}}

\begin{spacing}{1.6}
\noindent\textbf{\fontsize{10}{10}\selectfont{6.2.~~OSML Effectiveness}}
\end{spacing}
\noindent We compare OSML with the most related approaches in [10,46] based on the latest open-source version.

\noindent\textbf{PARTIES} [10]. It makes incremental adjustments in one-dimension resource at a time until QoS is satisfied – "trial and error" – for all of the applications. The core mechanism in [10] is like an FSM [60]. 

\noindent\textbf{CLITE} [46]. It conducts various allocation policies and samples each of them; it then feeds the sampling results – the QoS and run-time parameters for resources – to a Bayesian optimizer to predict the next scheduling policy. 

\noindent\textbf{Unmanaged Allocation (baseline)}. This policy doesn’t control the allocation policies on cores, LLC, and other shared resources for co-located LC services. This policy relies on the original OS schedulers.

\noindent\textbf{ORACLE.} We obtain these results by exhaustive offline sampling and find the best allocation policy. It indicates the ceiling that the schedulers try to achieve.

We show the effectiveness of OSML as follow. 

(1) OSML achieves a high EMU with shorter scheduling convergence time in most cases. Using ML models, OSML achieves OAA quickly and can efficiently handle cases with diverse loads. We tested 104 different loads for OSML, PARTIES and CLITE, respectively. Figure \ref{fig8}-a shows the distributions of the scheduling results for these 312 (104*3) cases. Every dot represents a scheduling case for a specific workload that contains several co-located LC services with diverse RPS. The x-axis shows the convergence time; the y-axis denotes the achieved EMU. Generally, OSML can achieve the same EMU with a shorter convergence time for a specific load. Figure \ref{fig8}-b shows the violin plots of convergence time for these loads. On average, OSML takes 20.9 seconds to converge, while PARTIES and CLITE take 32.7 and 46.3 seconds, respectively. OSML converges 1.56X and 2.22X faster than PARTIES and CLITE. OSML performs stably – the convergence time ranges from 5.3s (best case) to 80.0s (worst case). By contrast, the convergence time in PARTIES ranges from 5.5s to 111.1s, and CLITE is from 14.0s to 140.6s. OSML converges faster mainly because the start point in the scheduling space provided by Model-A is close to OAA. PARTIES and CLITE take a longer convergence time, indicating that they require high scheduling overheads in cloud environments. In Cloud, jobs come and go frequently; thus, scheduling occurs frequently, and longer scheduling convergence time often leads to unstable/low QoS.

\begin{figure}[!t]
\centering
\includegraphics[width=\linewidth]{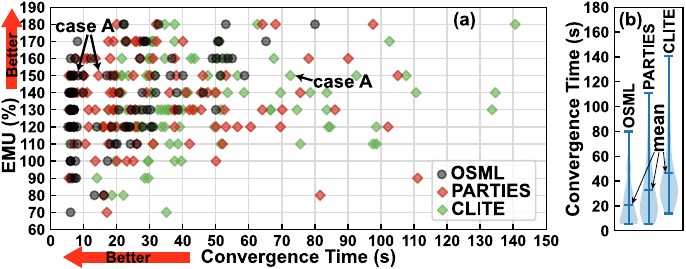}
\caption{\textcolor{black}{(a) The performance distributions for OSML, PARTIES, and CLITE; 104 different loads are tested for every scheduler. (b) Violin plots of convergence time for loads in (a).}} 
\label{fig8}
\end{figure}

\textcolor{black}{We further analyze how these schedulers work in detail. Figure \ref{fig9}-a/b/c show the actions used in OSML, PARTIES, and CLITE's scheduling process for case A in Figure \ref{fig8}. This case includes Moses, Img-dnn, and Xapian with 40\%, 60\%, and 50\% of their maximum loads. For this load, PARTIES, CLITE and OSML take 14.5 seconds, 72.6 seconds and 8.2 seconds to converge, respectively.} Figure \ref{fig9} highlights scheduling actions using solid red lines to represent increasing resources and blue dotted lines to denote reducing resources. \textcolor{black}{Figure \ref{fig9}-a shows PARTIES takes 7 actions for scheduling cores and 1 action for cache ways. It schedules in a fine-grained way by increasing/decreasing one resource at a time.} CLITE relies on the sampling points in the scheduling exploration space. \textcolor{black}{Figure \ref{fig9}-b shows CLITE repeats sampling until the ``expected improvement'' in CLITE drops below a certain threshold. CLITE only performs five scheduling actions according to its latest open-source version; but it takes the longest convergence time (72.6 seconds). The underlying reason is that CLITE's sampling/scheduling doesn't have clear targets. In practice, the improper resource partitions/allocations during sampling lead to the accumulation of requests, and the requests cannot be handled due to resource under-provision. Therefore, it brings a significant increase in response latency. Moreover, due to the early termination of CLITE's scheduling process, CLITE cannot schedule resources to handle QoS violations in a timely manner, leading to a long convergence time. Figure \ref{fig9}-c shows OSML achieves OAA for each LC service with 5 actions. Compared with prior schedulers, OSML has clear aims and schedules multiple resources simultaneously to achieve them. It has the shortest convergence time – 8.2 seconds.}

Moreover, as the scheduling is fast, OSML often supports more loads. \textcolor{black}{Figure \ref{fig10} shows the OSML’s results on scheduling the three LC services – Moses, Img-dnn, and Xapian. For a specific scheduling phase, by using ML to achieve OAA, OSML supports 10\textasciitilde50\% higher percentage of loads than PARTIES and CLITE (e.g., highlighted cells in Figure \ref{fig10}-d).} All schedulers perform better than the Unmanaged (Figure \ref{fig10}-a), as they reduce the resource contentions.

\begin{figure*}[tp]
	\flushleft
	\begin{minipage}[t!]{0.99\linewidth}
		\flushright
		\includegraphics[width=0.985\linewidth]{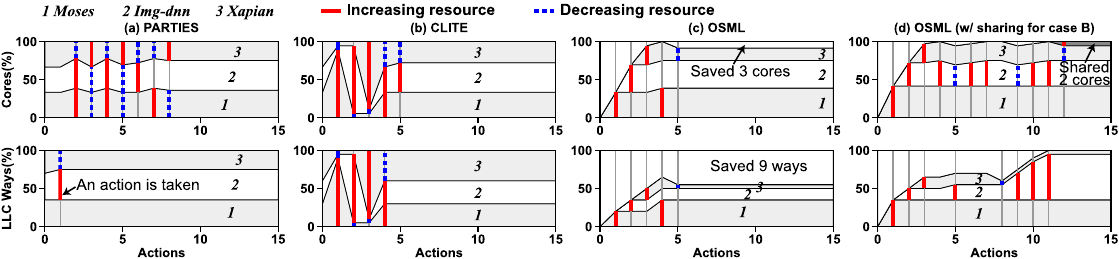}
		\caption{Resource usage comparisons for OSML, PARTIES, and CLITE.}
		\label{fig9}
	\end{minipage}

	\begin{minipage}[b!]{0.695\linewidth}
	\vspace{0.2cm}
	\centering
	\includegraphics[width=0.95\linewidth]{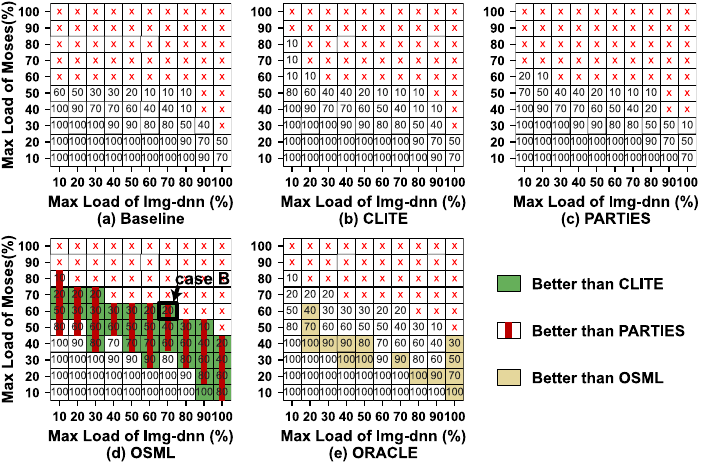}
	\vspace{0.5pt}
	\caption{\textls[-3]{Co-location of Moses, Img-dnn and Xapian. The heatmap values are the percentage of the third LC service’s (i.e., Xapian) achieved max load without QoS violations in these cases.} The x- and y-axis denote the first and second app’s fraction of their max loads (with QoS target), respectively. Cross means QoS target cannot be met. \textcolor{black}{The related studies [10,46] use heat maps to show their effectiveness, so we also use heat maps for comparisons in this work.}}
	\label{fig10}
	\end{minipage}
	\begin{minipage}[t!]{0.29\linewidth}
		\flushleft
		\includegraphics[width=0.97\linewidth]{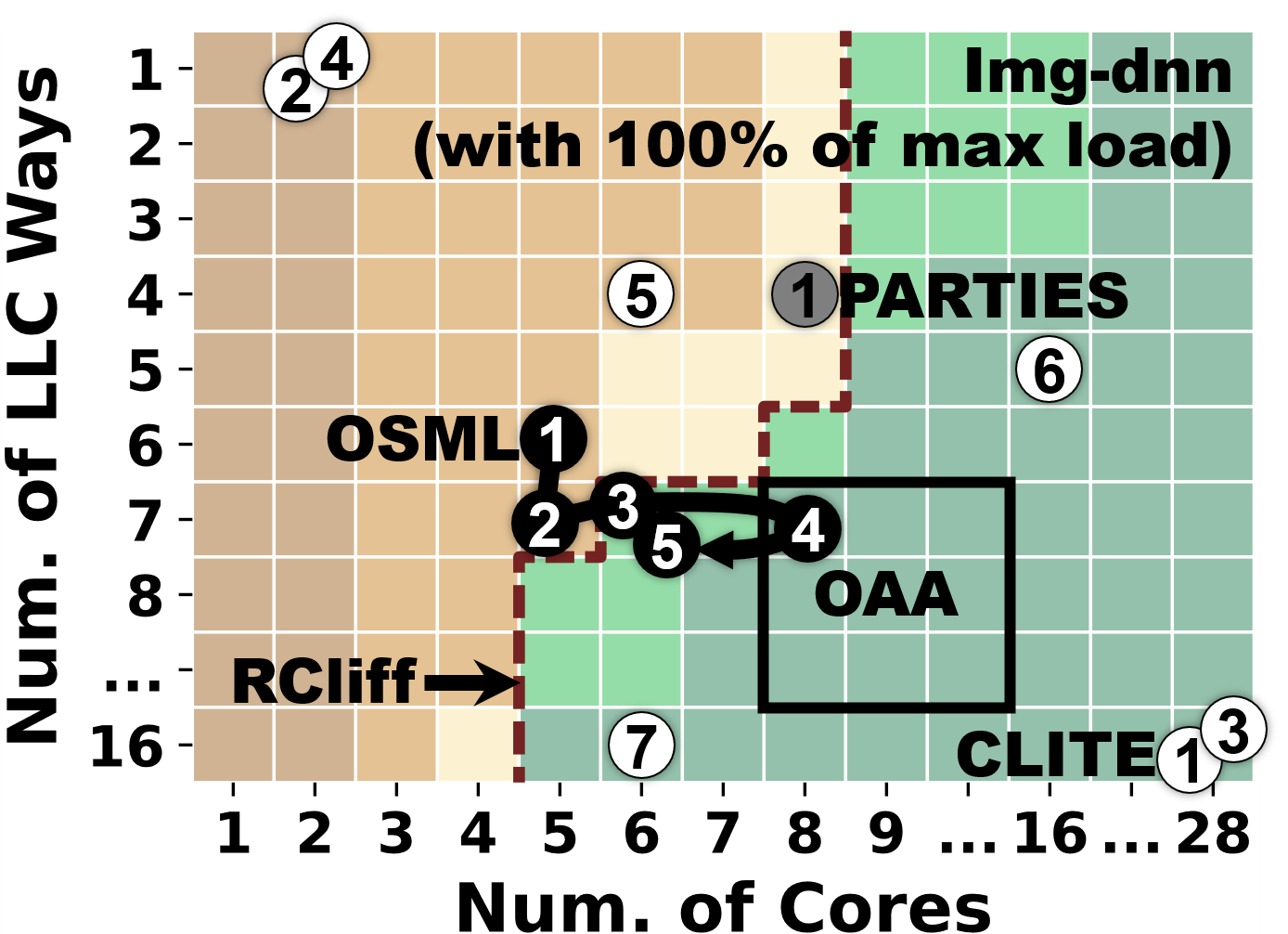}
		\setcounter{figure}{11}
		\vspace{-9pt}
		\caption{\textcolor{black}{Highlighted the scheduling traces in scheduling space for all schedulers from time point 180 to 228 in Figure 11. Each circle denotes a specific scheduling policy conducted by a specific scheduler. The number in a circle denotes the sequence of these scheduling actions during the scheduling phase.}}
		\label{fig11}
		\vspace{55pt}
	\end{minipage}
\end{figure*}

(2) Compared with PARTIES and CLITE, OSML uses fewer resources to support identical loads to meet the QoS targets. As illustrated in Figure \ref{fig9}-a, PARTIES partitions the LLC ways and cores equally for each LC service at the beginning; once it meets the QoS target (using 8 actions), it stops. Thus, PARTIES drops the opportunities to explore alternative better solutions (i.e., using fewer cores or cache ways to meet identical QoS targets). PARTIES allocates all cores and LLC ways finally. CLITE also uses all cores and cache ways shown in Figure \ref{fig9}-b. \textcolor{black}{By contrast, OSML schedules according to applications' resource requirements instead of using all resources. Figure \ref{fig9}-c shows that using Model-A, OSML achieves each LC service’s OAA (the optimal solution) after 5 actions. OSML detects and reclaims over-provided resources using Model-C. For example, the last action in Figure \ref{fig9}-c reclaims 3 cores and 2 LLC ways from Xapian. Finally, OSML saves 3 cores and 9 LLC ways.} As OSML is designed for LC services that are executed for a long period, saving resources means saving budgets for cloud providers.

\textheight=650pt

\textls[-3]{(3) Using ML models, OSML provides solutions for sharing some cores and LLC ways among LC services, therefore supporting higher loads. PARTIES and CLITE don't show resource sharing in the original design. Using Algo.\_4, OSML lists some potential resource sharing solutions, and then enables Model-B’ to predict the QoS slowdown for each case. The sharing solution with a relatively lower QoS slowdown is selected. More details refer to Figure \ref{fig7}. Figure \ref{fig9}-d shows how OSML shares resources for the highlighted case B in Figure \ref{fig10}-d. \textcolor{black}{OSML enables Model-C to add resources for Moses in Algo.\_2 and uses Algo.\_4 to share 2 CPU cores with Xapian.} Finally, the QoS is met. \textcolor{black}{By enabling resource sharing, OSML can support higher loads than PARTIES and CLITE, and can even be close to ORACLE in Figure \ref{fig10}-e.} If not OSML, however, the ``trial and error'' approach has to try to share core/cache way in a fine-grain way among applications, and then observes the real-time latency for making a further decision, inevitably incurring higher scheduling overheads and bringing sharp QoS slowdown if falling off the RCliff.}

(4) OSML promptly handles the resource under/over-provision and QoS violations using Model-C. Based on Model-A/B's results, Model-C shepherds and adjusts the allocations with several actions for each application in our experiments and converges more quickly than previous approaches. More experiments on dynamic, complicated cases can be found in Sec.6.3. \emph{Can we only use Model-C or only use Model-A/B?} Enabling the three models is necessary for OSML. For the case in Figure \ref{fig9}-c, when OSML uses the three models collaboratively for scheduling, it takes 8.2s and 5 actions to achieve OAA for all applications. By contrast, if only enabling Model-C, it takes 18.5s and 13 actions. Because Model-A/B can at least provide an approximate OAA for scheduling, thus reducing the convergence time. Just using Model-A/B may lead to 4-core errors for an unseen LC service (Table \ref{tbl5}). So, Model-C is needed to correct the errors. We cannot disable any models in OSML in practice.

\begin{figure}[!t]
\centering
\vspace{0.1cm}
\setcounter{figure}{10}
\includegraphics[width=0.99\linewidth]{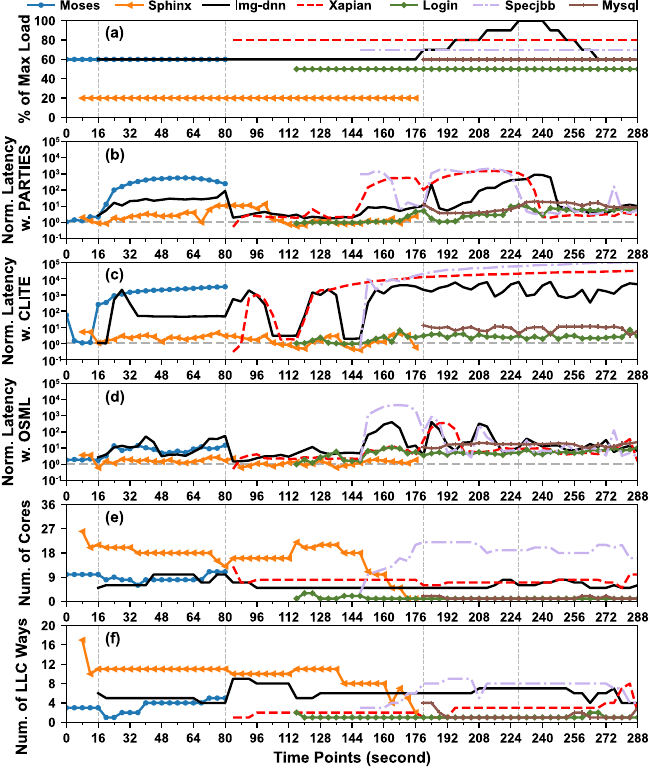}
\vspace{-1.3pt}
\caption{OSML's performance in reality with varying loads.}
\label{fig12}
\end{figure}

\begin{spacing}{1.6}
\noindent\textbf{\fontsize{10}{10}\selectfont{6.3.~~Performance for Workload Churn}}
\end{spacing}

\noindent We evaluate how OSML behaves with dynamically changing loads. Each LC service's QoS is normalized to the solely running case. As illustrated in Figure \ref{fig12}-a, in the beginning, Moses with 60\% of max load arrives; then Sphinx with 20\% of max load and Img-dnn with 60\% of max load arrive. We observe their response latency increases caused by the resource contentions among them. In Figure \ref{fig12}-b, PARTIES aids the QoS violations step by step. During the scheduling, Moses always has high latency until it ends at 80 seconds. CLITE's scheduling relies on sampling and Bayesian optimizer. CLITE starts scheduling at time point 16, where all the three services arrive. At 32s, CLITE obtains the scheduling solution for these LC services after five sampling steps. However, it does not meet Moses and Img-dnn's QoS targets. In Figure \ref{fig12}-c, Moses and Img-dnn still have high latency. By contrast, with Model-A's OAA predictions and Model-C's online scheduling, OSML quickly provides better scheduling solutions at time point 48 for all three services. During the identical scheduling phase (e.g., the time point 16 to 80), we can observe the lowest overall normalized latency in Figure \ref{fig12}-d. Moreover, Figure \ref{fig12}-e and f illustrate OSML’s scheduling actions for achieving ideal solutions. In short, within a few scheduling actions (scheduling overheads) that schedule multiple resources, OSML quickly meets the QoS targets.

From 180 to 228, we increase the load for Img-dnn as illustrated in Figure \ref{fig12}-a. OSML meets Img-dnn’s changing demands by using Model-C. PARTIES does not reflect quickly for this change, and it works for other services. Thus, as illustrated in Figure \ref{fig12}-b, the QoS violation is not aided until 244s, when Img-dnn's load decreases. For CLITE, it has to sample each time when the load changes. But during the sampling, a specific service might not have sufficient resources to handle the requests; thus, the requests are accumulated, leading to QoS fluctuations/violations during the scheduling (Figure \ref{fig12}-c). Figure \ref{fig11} highlights the scheduling actions for Img-dnn from 180 to 228. During this phase, PARTIES does not add resources for Img-dnn; but it add more resources for Specjbb and Xapian as they are with higher latency. Img-dnn's response latency keeps increasing. CLITE samples several scheduling policies in the scheduling space, but does not converge and thus incurs QoS fluctuations. By contrast, OSML's Model-C achieves Img-dnn's OAA using fewer scheduling actions. Moreover, as mentioned before, OSML saves resources ant thus it can serve more workloads. For example, as shown in Figure \ref{fig12}, Mysql (an \textbf{unseen} workload in training) comes at time point 180; OSML allocates the saved cores to it without sharing or depriving other LC services of resources.

\begin{scriptsize}
\begin{table}[t]\small
  \setlength{\tabcolsep}{0.9pt}
  \centering
  \setlength{\extrarowheight}{1pt}
\caption{ML Models’ performance on average.}
\begin{tabular}{|p{0.05\linewidth}<{\centering}| p{0.20\linewidth}<{\centering} | p{0.08\linewidth}<{\centering} |p{0.08\linewidth}<{\centering}| p{0.08\linewidth}<{\centering} | p{0.08\linewidth}<{\centering} |p{0.08\linewidth}<{\centering}| p{0.08\linewidth}<{\centering} | p{0.09\linewidth}<{\centering} | p{0.08\linewidth}<{\centering} | p{0.14\linewidth}<{\centering} |}
    \hline
    \multirow{2}*{\makecell[c]{\\[-1.5pt]\textbf{ML}}}
    &\multirow{2}*{\makecell[c]{\\[-1.5pt]\textbf{Outputs}}}
    &\multicolumn{2}{c|}{\makecell[c]{\\[-3.5pt]\textbf{Error}}}
    &\multicolumn{2}{m{0.16\linewidth}<{\centering}|}{
    \makecell[c]{\\[-8pt]\textbf{\textcolor{black}{Errors for}}\\[-2pt]\textbf{\textcolor{black}{unseen LC}}\\[-2pt]\textbf{\textcolor{black}{services}}\\[-1pt]}}
    &\multicolumn{2}{m{0.16\linewidth}<{\centering}|}{ \makecell[c]{
    \\[-8pt]\textbf{Err on new}\\[-2pt]\textbf{platforms}\\[-2pt]\textbf{(TL)}\\[-1pt]}}
    &\multirow{2}*{\makecell[c]{\\[-1.5pt]\textbf{MSE}}}
    &\multirow{2}*{\makecell[c]{\\[-1.5pt]\textbf{Over-}\\[-1.5pt]\textbf{heads}}}\\
    \cline{3-8}
    ~&~&\textbf{Core}&\textbf{LLC}&\textbf{Core}&\textbf{LLC}&\textbf{Core}&\textbf{LLC}&~&~\\
    \hline
    \hline
	\multirow{2}*{A}&	RCliff&	0.589&	0.288&	\textcolor{black}{1.266}&	\textcolor{black}{0.198}&  \textcolor{black}{2.142}&  \textcolor{black}{0.542}&	\multirow{2}*{0.0025}&	\multirow{2}*{0.20s}\\
	\cline{2-8}
	~				&	OAA&	0.580&	0.360&	\textcolor{black}{1.276}&	\textcolor{black}{0.191}&  \textcolor{black}{2.004}&  \textcolor{black}{0.865}&	~&~	\\
	\hline
	\multirow{2}*{A’}&	RCliff&	1.072&	0.815&	3.403&	1.835&\textcolor{black}{0.772}&\textcolor{black}{0.411}&	\multirow{2}*{0.0135}&\multirow{2}*{0.20s}\\
	\cline{2-8}
	~				&	OAA&	1.072&	0.814&	3.404&	1.835&\textcolor{black}{0.790}&\textcolor{black}{0.413}&~&~	\\
	\hline
	\multirow{5}*{B}&	B-Points&	0.612&	0.053&	4.012&	0.167&\textcolor{black}{2.320}&\textcolor{black}{0.969}& \multirow{5}*{0.0012}&\multirow{5}*{0.18s}\\
	\cline{2-8}
	~				&	\multicolumn{1}{m{0.20\linewidth}<{\centering}|}{\fontsize{7.5}{5} \selectfont B-Points,Core dominated}&	0.314&	0.048&3.434&0.937&\textcolor{black}{2.250}&\textcolor{black}{0.815}& ~&~		\\
	\cline{2-8}
	~				&	\multicolumn{1}{m{0.20\linewidth}<{\centering}|}{\fontsize{7.5}{5} \selectfont B-Points,Cache dominated}&	0.093&	0.462&0.789&0.783&\textcolor{black}{1.868}&\textcolor{black}{1.519}&	~&~	\\
	\hline
	B’&	QoS reduction&	\multicolumn{2}{c|}{7.87\%}&\multicolumn{2}{c|}{8.33\%}&\multicolumn{2}{c|}{\textcolor{black}{11.28\%}}&	0.0035&	0.19s\\
	\hline
	C&
	\multicolumn{1}{m{0.20\linewidth}<{\centering}|}{\fontsize{7.5}{5} \selectfont Scheduling actions}&0.908&0.782&0.844&0.841&\textcolor{black}{1.390}&\textcolor{black}{1.801}&0.7051&0.20s\\
	\hline
  \end{tabular}
  \label{tbl5}
  \vspace{-3pt}
\end{table}
\end{scriptsize}

\begin{spacing}{1.6}
\noindent\textbf{\fontsize{10}{10}\selectfont{6.4.~~Evaluations for New/Unseen Apps and New Platforms}}
\end{spacing}
\noindent \textbf{Generalization.} Based on our comprehensive data set, the ML models are well trained; and it is possible to skip profiling for new/unseen applications. The sensitivity will be at most 4-core error (Table \ref{tbl5}). We evaluate OSML for unseen applications that are not in Table \ref{tbl1}, e.g., Silo [62], Shore [62], and Mysql [70]. OSML provides high-quality predictions (i.e., the errors for OAAs and RCliffs are less than one core and one LLC way) for 52.0\% of all cases. 89.4\% of all cases have errors less than three cores and three ways, on average. Model-C can correct these errors within one or two actions on the fly. For scheduling time, we construct ten concurrent workloads by mixing these new ones with the applications in Table \ref{tbl1}. The workloads contain at least one new application. The experimental results show that OSML, PARTIES, and CLITE take 29.7s, 39.2s, and 64.8s to schedule these workloads, on average, respectively.

\textls[-5]{\textbf{For new platforms}, we use fine-tuning in transfer learning (TL). We freeze the first hidden layer of the MLPs; we retrain the last two-hidden layers and the output layer using the traces collected on \textcolor{black}{two new platforms (w/ CPU Xeon Gold 6240M and E5-2630 v4, respectively)}. For each LC service, based on our data set, collecting new traces on a new platform for several hours will be sufficient (covering the more allocation cases, the better). The time consumption will be shorter if using multiple machines in parallel. \textcolor{black}{Table \ref{tbl5} shows the average values of ML models' quality. The new models' prediction errors are slightly higher than the previous models on the original platforms, but OSML still handles them well.} By contrast, if we use a \textcolor{black}{\textbf{table lookup approach}} instead, we have to use additional memory to store the data tuples, e.g., 60GB will be wasted for the current data set to replace Model-A. \textcolor{black}{More importantly, it is difficult to generalize a table look-up approach for new/unseen applications or platforms, as their traces and the corresponding OAAs don't exist in the current data set.}}

\textbf{Overheads.} OSML takes 0.2s for each time (\textcolor{black}{0.01s for ML model and 0.19s for online monitoring}). As our models are light-weighted (\textcolor{black}{OSML uses only one core}), running them on CPU and GPU has a similar overhead. If models are on GPU, it takes an extra 0.03s for receiving results from GPU. OSML doesn't monopolize GPU. Generally, the overhead doesn't hinder the overall performance. In the cloud, applications' behaviors may change every second due to the diversity of user demands. Thus, OSML plays a critical role during the entire run time. \textbf{For training time}, using our current data set, it takes 3.3 mins, 5 mins, and 8.3 hours to train Model-A, B, and C for one epoch, respectively. One epoch means that all training samples in the data set are used to train once. We train models for ten epochs. Training can be accelerated using the popular Multi-GPU training technology - using multiple GPUs simultaneously to train one model. Doing so is practical in datacenters, and training time will not impede practice.\\[-1pt]

\noindent\textbf{\fontsize{12px}{12px}\selectfont{7.~~Related Work and Our Novelty}}
\\[-3pt]

\noindent \textbf{ML for System Optimizations.} The work in [55] employs DNN to optimize the buffer size for the database systems. The efforts in [22,56,34] leverage ML to optimize computer architecture or resource management in the network to meet various workloads. The studies in [9,39] use ML to manage on-chip hardware resources. CALOREE [41] can learn key control parameters to meet latency requirements with minimal energy in complex environments. The studies in [26,31,58,59] optimize the OS components with learned rules or propose insights on designing new learned OS. \emph{In OSML, we design an intelligent multi-model collaborative learning approach, providing better co-location solutions to meet QoS targets for LC services faster than the latest work stably.}

\noindent \textbf{ML for Scheduling.} Decima [35] designs cluster-level data processing job scheduling using RL. Resource Central [12] builds a system that contains the historical resource utilization information of the workloads used in Azure and employs ML to predict resource management for VMs. [40] uses RL to predict which subsets of operations in a TensorFlow graph should run on the available devices. Paragon [14] classifies and learns workload interference. Quasar [15] determines jobs' resource preferences on clusters. Sinan [74] uses ML models to determine the performance dependencies between microservices in clusters. They are cluster schedulers [14,15,74]. By contrast, OSML deeply studies scheduling in co-location cases. Selecta [72] predicts near-optimal configurations of computing and storage resources for analytics workloads based on profiling data. CLITE [46] uses Bayesian optimization for scheduling on-node resources. The work in [48] applies ML to predict the end-to-end tail latency of LC service workflows. Twig [63] uses RL to characterize tail latency for energy-efficient task management. \textcolor{black}{CuttleSys [76] leverages data mining to identify suitable core and cache configurations for co-scheduled applications.} \emph{For complicated co-location cases on a specific cloud server, using the fewest scheduling actions on average compared with the latest studies, OSML can avoid RCliff and achieve the ideal allocations (OAA) for multiple interactive resources simultaneously for LC services. Moreover, OSML performs well in generalization.}

\noindent\textbf{Resource Partitioning.} PARTIES [10] partitions cache, main memory, I/O, network, disk bandwidth, etc., to provide QoS for co-located services. The studies in [17,28,57,71] design some new LLC partitioning/sharing policies. The efforts in [23,27,44,45,73] show that considering cooperative partitioning on LLC, memory banks and channels outperforms one-level memory partitioning. However, the cooperative partitioning policies need to be carefully designed [29,30,37], and [16,32] show the heuristic resource scheduling approach could be ineffective in many QoS-constrained cases. [7,11] study the “performance cliff” on cache for SPECCPU 2006 applications and Memcached. \textcolor{black}{Caladan [75] doesn't involve cache optimizations, and core/cache cliffs cannot be avoided, causing QoS fluctuations in some cases.} \emph{By contrast, OSML is the first work that profoundly explores cache cliff and core cliff simultaneously (i.e., RCliff) for many widely used LC services in co-location cases. OSML is a representative work using ML to guide the multiple resources partitioning in co-location cases; OSML is cost-effective in new cloud environments.}\\[-1pt]

\noindent\textbf{\fontsize{12px}{12px}\selectfont{8.~~Conclusion}}
\\[-3pt]

\noindent \textcolor{black}{We present OSML, an resource scheduler for LC services. OSML employs ML to preserve QoS for the co-scheduled services. We should OSML performs well. We also learn that straightforwardly using a simple ML model might not handle all of the processes during the scheduling. Therefore, using multiple ML models cooperatively in a pipe-lined way can be an ideal approach. More importantly, we advocate the new solution, i.e., leveraging ML to enhance resource scheduling, could have an immense potential for OS design. In a world where co-location and sharing are a fundamental reality, our solution should grow in importance and benefits our community.}


\begin{thebibliography}{10}

\providecommand{\justify}{\leftskip=0pt \rightskip=0pt plus 0cm}

\justify

\bibitem{ref1}
“How 1s could cost amazon \$1.6 billion in sales.” https://www.fastcompany.com/1825005/how-one-second-could-cost-amazon-16-billion-sales

\bibitem{ref2}
“Microservices workshop: Why, what, and how to get there,” http://www.slideshare.net/adriancockcroft/microservices-workshop-craft-conference

\bibitem{ref3}
“State of the Cloud Report,” http://www.righscale.com/lp/state-of-the-cloud. Accessed: 2019-01-28

\bibitem{ref4}
“Improving real-time performance by utilizing cache allocation technology,” https://www.intel.com/content/dam/www/public/us/en/documents/white-papers/cache-allocation-technology-white-paper.pdf, Intel Corporation, April, 2015

\bibitem{ref5}
“Intel 64 and IA-32 Architectures Software Developer’s Manual,” https://software.intel.com/en-us/articles/intel-sdm, Intel Corporation, October, 2016

\bibitem{ref6}
Martín Abadi, Paul Barham, Jianmin Chen, Zhifeng Chen, Andy Davis, Jeffrey Dean, Matthieu Devin, Sanjay Ghemawat, Geoffrey Irving, Michael Isard, Manjunath Kudlur,  Josh Levenberg, Rajat Monga, Sherry Moore, Derek G. Murray, Benoit Steiner, Paul Tucker, Vijay Vasudevan, Pete Warden, Martin Wicke, Yuan Yu, and Xiaoqiang Zheng, “TensorFlow: A System for Large-Scale  Machine Learning,” in OSDI, 2016

\bibitem{ref7}
Nathan Beckmann, Daniel Sanchez, “Talus: A Simple Way to Remove Cliffs in Cache Performance,” in HPCA, 2015

\bibitem{ref8}
Daniel S. Berger, Benjamin Berg, Timothy Zhu, Siddhartha Sen, Mor Harchol-Balter, “RobinHood: Tail Latency Aware Caching -- Dynamic Reallocation from Cache-Rich to Cache-Poor,” in OSDI, 2018

\bibitem{ref9}
Ramazan Bitirgen, Engin Ipek, Jose F. Martinez, “Coordinated Management of Multiple Interacting Resources in Chip Multiprocessors: A Machine Learning Approach,” in Micro, 2008

\bibitem{ref10}
Shuang Chen, Christina Delimitrou, José F. Martínez, “PARTIES: QoS-Aware Resource Partitioning for Multiple Interactive Services,” in ASPLOS, 2019

\bibitem{ref11}
Asaf Cidon, Assaf Eisenman, Mohammad Alizadeh, Sachin Katti, “Cliffhanger: Scaling Performance Cliffs in Web Memory Caches,” in NSDI, 2016

\bibitem{ref12}
Eli Cortez, Anand Bonde, Alexandre Muzio, Mark Russinovich, Marcus Fontoura, Ricardo Bianchini, “Resource Central: Understanding and Predicting Worloads for Improved Resource Management in Large Cloud Platforms,” in SOSP, 2017

\bibitem{ref13}
Jeff Dean, David A. Patterson, Cliff Young, “A New Golden Age in Computer Architecture: Empowering the Machine-Learning Revolution,” in IEEE Micro 38 (2): 21-29 (2018)

\bibitem{ref14}
Christina Delimitrou, Christos Kozyrakis, “QoS-Aware Scheduling in Heterogenous Datacenters with Paragon,” in ACM TOCS, 2013

\bibitem{ref15}
Christina Delimitrou, Christos Kozyrakis, “Quasar: Resource-Efficient and QoS-Aware Cluster Management,” in ASPLOS, 2014

\bibitem{ref16}
Yi Ding, Nikita Mishra, Henry Hoffmann, “Generative and Multi-phase Learning for Computer Systems Optimization,” in ISCA, 2019

\bibitem{ref17}
Nosayba El-Sayed, Anurag Mukkara, Po-An Tsai, Harshad Kasture, Xiaosong Ma, Daniel Sanchez, “KPart: A hybrid Cache Partitioning-Sharing Technique for Commodity Multicores,” in HPCA, 2018

\bibitem{ref18}
Yu Gan and Christina Delimitrou, “The Architectural Implications of Cloud Microservices,” in IEEE Computer Architecture Letters, 2018

\bibitem{ref19}
Yu Gan, Yanqi Zhang, Kelvin Hu, Dailun Cheng, Yuan He, Meghna Pancholi, Christina Delimitrou, “Leveraging Deep Learning to Improve Performance Predictability in Cloud Microservices with Seer,” in ACM SIGOPS Operating Systems Review, 2019

\bibitem{ref20}
Mark D. Hill, Michael R. Marty, “Amdahl's Law in the Multicore Era,” in IEEE Computers, 2008

\bibitem{ref21}
Kurt Hornik, “Approximation Capabilities of Multilayer Feedforward Networks,” in Neural Networks, 1991

\bibitem{ref22}
Engin Ipek, Onur Mutlu, José F. Martínez, Rich Caruana, “Self-Optimizing Memory Controllers: A Reinforcement Learning Approach,” in ISCA, 2008

\bibitem{ref23}
Min Kyu Jeong, Doe Hyun Yoon, Dam Sunwoo, Michael Sullivan, Ikhwan Lee, Mattan Erez,“Balancing DRAM Locality and Parallelism in Shared Memory CMP Systems,”in HPCA, 2012

\bibitem{ref24}
Norman P. Jouppi, Cliff Young, Nishant Patil, David Patterson, Gaurav Agrawal, Raminder Bajwa, Sarah Bates, Suresh Bhatia, Nan Boden, Al Borchers, Rick Boyle, Pierre-luc Cantin, Clifford Chao, Chris Clark, Jeremy Coriell, Mike Daley, Matt Dau, Jeffrey Dean, Ben Gelb, Tara Vazir Ghaemmaghami, Rajendra Gottipati, William Gulland, Robert Hagmann, C. Richard Ho, Doug Hogberg, John Hu, Robert Hundt, Dan Hurt, Julian Ibarz, Aaron Jaffey, Alek Jaworski, Alexander Kaplan, Harshit Khaitan, Daniel Killebrew, Andy Koch, Naveen Kumar, Steve Lacy, James Laudon, James Law, Diemthu Le, Chris Leary, Zhuyuan Liu, Kyle Lucke, Alan Lundin, Gordon MacKean, Adriana Maggiore, Maire Mahony, Kieran Miller, Rahul Nagarajan, Ravi Narayanaswami, Ray Ni, Kathy Nix, Thomas Norrie, Mark Omernick, Narayana Penukonda, Andy Phelps, Jonathan Ross, Matt Ross, Amir Salek, Emad Samadiani, Chris Severn, Gregory Sizikov, Matthew Snelham, Jed Souter,  Dan Steinberg, Andy Swing, Mercedes Tan, Gregory Thorson, Bo Tian, Horia Toma, Erick Tuttle,  Vijay Vasudevan, Richard Walter, Walter Wang, Eric Wilcox, Doe Hyun Yoon, “In-Datacenter Performance Analysis of a Tensor Processing Unit,” in ISCA, 2017

\bibitem{ref25}
Alex Krizhevsky, Ilya Sutskever, Geoffrey Hinton, "ImageNet Classification with Deep Convolutional Neural Networks," in Advances in neural information processing systems, 2012

\bibitem{ref26}
Yanjing Li, Onur Mutlu, Subhasish Mitra, “Operating System Scheduling for Efficient Online Self-Test in Robust Systems,” in ICCAD, 2009

\bibitem{ref27}
Lei Liu, Zehan Cui, Mingjie Xing, Chengyong Wu, “A Software Memory Partition Approach for Eliminating Bank-level Interference in Multicore Systems,” in PACT, 2012

\bibitem{ref28}
Jiang Lin, Qingda Lu, Xiaoning Ding, Zhao Zhang, Xiaodong Zhang, P. Sadayappan, “Gaining insights into mlticore cache partitioning: bridging the gap between simulation and real systems,” in HPCA, 2008

\bibitem{ref29}
Fang Liu, Yan Solihin, “Studying the Impact of Hardware Prefetching and Bandwidth Partitioning in Chip-Multiprocessors,” in Sigmetrics, 2011

\bibitem{ref30}
Seung-Hwan Lim, Jae-Seok Huh, Yougjae Kim, Galen M. Shipman, Chita R. Das, “D-Factor: A Quantitative Model of Application Slow-Down in Multi-Resource Shared Systems,” in Sigmetrics, 2012

\bibitem{ref31}
Lei Liu, Yong Li, Chen Ding, Hao Yang, Chengyong Wu, “Rethinking Memory Management in Modern Operating System: Horizontal, Vertical or Random?” in IEEE Trans. on Computers, 2016

\bibitem{ref32}
David Lo, Liqun Cheng, Rama Govindaraju, Parthasarathy Ranganathan, Christos Kozyrakis, "Heracles: Improving Resource Efficiency at Scale," in ISCA, 2015

\bibitem{ref33}
Lei Liu, Shengjie Yang, Lu Peng, Xinyu Li, “Hierarchical Hybrid Memory Management in OS for Tiered Memory Systems,” in IEEE Trans. on Parallel and Distributed Systems, 2019

\bibitem{ref34}
Hongzi Mao, Mohammad Alizadeh, Ishai Menache, Srikanth Kandula, “Resource Management with Deep Reinforcement Learning,” in HotNet-XV, 2016

\bibitem{ref35}
Hongzi Mao, Malte Schwarzkopf, Shaileshh B. Venkatakrishnan, Zili Memg, Mohammad Alizadeh, “Learning Scheduling Algorithms for Data Processing Clusters,” in SIGCOMM, 2019

\bibitem{ref36}
Yashwant Marathe, Nagendra Gulur, Jee Ho Ryoo, Shuang Song, and Lizy K. John, “CSALT: Context Switch Aware Large TLB,” in Micro, 2017

\bibitem{ref37}
Jason Mars, Lingjia Tang, Robert Hundt, Kevin Skadron, Mary Lou Soffa, “Bubble-Up: Increasing Utilization in Modern Warehouse Scale Computers via Sensible Co-locations,” in Micro, 2011

\bibitem{ref38}
Jason Mars, Lingjia Tang, Mary Lou Soffa, “Directly Characterizing Cross Core Interference Through Contention Synthesis,” in HiPEAC, 2011

\bibitem{ref39}
Jose F. Martinez, Egin Ipek, “Dynamic multicore resource management: A machine learning approach,” in IEEE Micro 29 (5):8-17 (2009)

\bibitem{ref40}
Azalia Mirhoseini, Hieu Pham, Quoc V. Le, Benoit Steiner, Rasmus Larsen, Yuefeng Zhou, Naveen Kumar, Mohammad Norouzi, Samy Bengio, Jeff Dean, "Learning Device Placement in Tensorflow Computations," in Arxiv 1706.04972

\bibitem{ref41}
Nikita Mishra, Connor Imes, John D. Lafferty, Henry Hoffmann, “CALOREE: Learning Control for Predictable Latency and Low Energy,” in ASPLOS, 2018

\bibitem{ref42}
Nikita Mishra, Harper Zhang, John Lafferty, Henry Hoffmann, “A probabilistic Graphical Model-based Approach for Minimizing Energy Under Performance Constraints,” in ASPLOS, 2015

\bibitem{ref43}
Volodymyr Mnih, Koray Kavukcuoglu, David Silver, Andrei A. Rusu, Joel Veness, Marc G. Bellemare, Alex Graves, Martin Riedmiller, Andreas K. Fidjeland, Georg Ostrovski, Stig Petersen, Charles Beattie, Amir Sadik, Ioannis Antonoglou, Helen King, Dharshan Kumaran, Daan Wierstra, Shane Legg, Demis Hassabis, “Human-level control through deep reinforcement learning,” in Nature 518 (7540): 529-533, 2015

\bibitem{ref44}
Sai Prashanth Muralidhara, Lavanya Subramanian, Onur Mutlu, Mahmut Kandemir, Thomas Moscibroda, “Reducing Memory Interference in Multicore Systems via Application-Aware Memory Channel Partitioning,” in Micro, 2011

\bibitem{ref45}
Jinsu Park, Seongbeom Park, Woongki Baek, “CoPart: Coordinated Partitioning of Last-Level Cache and Memory Bandwidth for Fairness-Aware Workload Consolidation on Commodity Servers,” in EuroSys, 2019

\bibitem{ref46}
Tirthak Patel, Devesh Tiwari, “CLITE: Efficient and QoS-Aware Co-Location of Multiple Latency-Critical Jobs for Warehouse Scale Computers,” in HPCA, 2020

\bibitem{ref47}
Henry Qin, Qian Li, Jacqueline Speiser, Peter Kraft, and John Ousterhout, “Arachne: Core-Aware Thread Management,” in OSDI, 2018

\bibitem{ref48}
Joy Rahman, Palden Lama, “Predicting the End-to-End Tail Latency of Containerized Microservices in the Cloud,” in IC2E, 2019

\bibitem{ref49}
Yizhou Shan, Yutong Huang, Yilun Chen, Yiying Zhang, “LegoOS: A Disseminated, Distributed OS for Hardware Resource Disaggregation,” in OSDI, 2018

\bibitem{ref50}
Prateek Sharma, Ahmed Ali-Eldin, Prashant Shenoy, “Resource Deflation: A New Approach For Transient Resource Reclamation,” in EuroSys, 2019

\bibitem{ref51}
David Silver, Aja Huang, Chris J. Maddison, Arthur Guez, Laurent Sifre, George van den Driessche,  Julian Schrittwieser, Ioannis Antonoglou, Veda Panneershelvam, Marc Lanctot, Sander Dieleman, Dominik Grewe, John Nham, Nal Kalchbrenner, Ilya Sutskever, Timothy Lillicrap, Madeleine Leach, Koray Kavukcuoglu, Thore Graepel, Demis Hassabis, “Mastering the game of Go with deep neural networks and tree search,” in Nature, 529 (7587), 2016

\bibitem{ref52}
Akshitha Sriraman, Abhishek Dhanotia, Thomas F. Wenisch, “SoftSKU: Optimizing Server Architectures for Microservice Diversity @Scale,” in ISCA, 2019

\bibitem{ref53}
Akshitha Sriraman, Thomas F. Wenisch, “µTune: Auto-Tuned Threading for OLDI Microservices,” in OSDI, 2018

\bibitem{ref54}
Christian Szegedy, Wei Liu, Yangqing Jia, Pierre Sermanet, Scoott Reed, Dragomir Anguelov, Dumitru Erhan, Vincent Vanhoucke, Andrew Rabinovich, “Going deeper with convolutions,” in CVPR, 2015

\bibitem{ref55}
Jian Tan, Tieying Zhang, Feifei Li, Jie Chen, Qixing Zheng, Ping Zhang, Honglin Qiao, Yue Shi, Wei Cao, Rui Zhang, “iBTune: Individualized Buffer Tuning for Large-scale Cloud Databases,” in VLDB, 2019

\bibitem{ref56}
Stephen J. Tarsa, Rangeen Basu Roy Chowdhury, Julien Sebot, Gautham Chinya, Jayesh Gaur, Karthik Sankaranarayanan, Chit-Kwan Lin, Robert Chappell, Ronak Singhal, Hong Wang, “Post-Silicon CPU Adaptations Made Practical Using Machine Learning,” in ISCA, 2019

\bibitem{ref57}
Xiaodong Wang, Shuang Chen, Jeff Setter, Jose F. Martínez, “SWAP: Effective Fine-Grain Management of Shared Last-Level Caches with Minimum Hardware Support,” in HPCA, 2017

\bibitem{ref58}
Zi Yan, Daniel Lustig, David Nellans, and Abhishek Bhattacharjee, “Nimble Page Management for Tiered Memory Systems,” in ASPLOS, 2019

\bibitem{ref59}
Yiying Zhang, Yutong Huang, “Learned Operating Systems,” in ACM SIGOPS Operating Systems Review, 2019

\bibitem{ref60}
Zhijia Zhao, Bo Wu, Xipeng Shen, “Challenging the "Embarrassingly Sequential": Parallelizing Finite State Machine-based Computations through Principled Speculation,” in ASPLOS, 2014

\bibitem{ref61}
Xiaoya Xiang, Chen Ding, Hao Luo, Bin Bao, “HOTL: A higher order theory of locality,” in ASPLOS, 2013

\bibitem{ref62}
Harshad Kasture, Daniel Sanchez, “Tailbench: a benchmark suite and evaluation methodology for latency-critical applications,” in IISWC, 2016

\bibitem{ref63}
Rajiv Nishtala, Vinicius Petrucci, Paul Carpenter, Magnus Sjalander, “Twig: Multi-Agent Task Management for Colocated Latency-Critical Cloud Services,” in HPCA, 2020

\bibitem{ref64}
MongoDB official website. http://www.mongodb.com

\bibitem{ref65}
Memcached official website. https://memcached.org

\bibitem{ref66}
NGINX official website. http://nginx.org

\bibitem{ref67}
https://en.wikipedia.org/wiki/Universal\_approximation\_theorem

\bibitem{ref68}
Yu Gan, Yanqi Zhang, Dailun Cheng, Ankitha Shetty, Priyal Rathi, Nayan Katarki, Ariana Bruno, Justin Hu, Brian Ritchken, Brendon Jackson, Kelvin Hu, Meghna Pancholi, Yuan He, Brett Clancy, Chris Colen, Fukang Wen, Catherine Leung, Siyuan Wang, Leon Zaruvinsky, Mateo Espinosa, Rick Lin, Zhongling Liu, Jake Padilla, Christina Delimitrou, “An Open-Source Benchmark Suite for Microservices and Their Hardware-Software Implications for Cloud \& Edge Systems,” in ASPLOS, 2019

\bibitem{ref69}
Kalyanmoy Deb, Shivam Gupta, “Understanding Knee Points in Bicriteria Problems and Their Implications as Preferred Solution Principles,” in Engineering Optimization, 43 (11), 2011

\bibitem{ref70}
www.mysql.com

\bibitem{ref71}
Harshad Kasture, Daniel Sanchez, “Ubik: Efficient Cache Sharing with Strict QoS for Latency-Critical Workloads,” in ASPLOS, 2014

\bibitem{ref72}
Ana Klimovic, Heiner Litz, Christos Kozyrakis, “Selecta: Learning Heterogeneous Cloud Storage Configuration for Data Analytics,” in USENIX ATC, 2018

\bibitem{ref73}
Harshad Kasture, Xu Ji, Nosayba El-Sayed, Xiaosong Ma, Daniel Sanchez, “Improving Datacenter Efficiency Through Partitioning-Aware Scheduling,” in PACT, 2017

\bibitem{ref74}
Yanqi Zhang, Weizhe Hua, Zhuangzhuang Zhou, G. Edward Suh, Christina Delimitrou, “Sinan: ML-Based and QoS-Aware Resource Management for Cloud Microservices,” in ASPLOS, 2021

\bibitem{ref75}
Joshua Fried, Zhenyuan Ruan, Amy Ousterhout, Adam Belay, “Caladan: Mitigating Interference at Microsecond Timescales,” in OSDI, 2020

\bibitem{ref76}
Neeraj Kulkarni, Gonzalo Gonzalez-Pumariega, Amulya Khurana, Christine A Shoemaker, Christina Delimitrou, David H Albonesi, “Cuttlesys: Data-driven Resource Management for Interactive Services on Re configurable Multicores,” in Micro, 2020

\end{thebibliography}

\end{document}